\documentclass[journal]{IEEEtran}
\usepackage{dsfont}
\usepackage{amssymb}
\usepackage{amsmath}
\usepackage{cite}
\usepackage{graphicx}
\usepackage{multicol}
\usepackage{caption}
\usepackage{subfigure}
\usepackage{booktabs}
\usepackage{nccmath}
\usepackage{setspace}
\usepackage{epstopdf}
\usepackage{fancyhdr}

\author{\IEEEauthorblockN{Yun Wei$^1$, Chuanyi Ji$^1$, Floyd Galvan$^2$, Stephen Couvillon$^2$, George Orellana$^2$, James Momoh$^3$} \\
\IEEEauthorblockA{$^1$Georgia Institute of Technology, Atlanta, GA 30332--0250 \\
$^2$Entergy Services, Inc., New Orleans, LA 70053 \\
$^3$Howard University, NW Washington, DC, 20059 \\
Email: \{yunwei, jichuanyi\}@gatech.edu}}
\title{Learning Geo-Temporal Non-Stationary Failure and Recovery of Power Distribution}

\begin{document}

\maketitle

\thispagestyle{fancy}
\fancyhead{}
\lhead{\bfseries Accepted with minor revisions by TNNLS \\
Special Issue on Learning in Nonstationary and Evolving Environments}
\chead{}
\rhead{}
\lfoot{}
\cfoot{\thepage}
\rfoot{}
\renewcommand{\headrulewidth}{0pt}
\renewcommand{\footrulewidth}{0.7pt}

\begin{abstract}
Smart energy grid is an emerging area for new applications of machine learning in a non-stationary environment. Such a non-stationary environment emerges when large-scale failures occur at power distribution networks due to external disturbances such as hurricanes and severe storms. Power distribution networks lie at the edge of the grid, and are especially vulnerable to external disruptions. Quantifiable approaches are lacking and needed to learn non-stationary behaviors of large-scale failure and recovery of power distribution. This work studies such non-stationary behaviors in three aspects. First, a novel formulation is derived for an entire life cycle of large-scale failure and recovery of power distribution. Second, spatial-temporal models of failure and recovery of power distribution are developed as geo-location based multivariate non-stationary $GI(t)/ G(t) / \infty$ queues. Third, the non-stationary spatial-temporal models identify a small number of parameters to be learned. Learning is applied to two real-life examples of large-scale disruptions. One is from Hurricane Ike, where data from an operational network is exact on failures and recoveries. The other is from Hurricane Sandy, where aggregated data is used for inferring failure and recovery processes at one of the impacted areas. Model parameters are learned using real data. Two findings emerge as results of learning: (a) Failure rates behave similarly at the two different provider networks for two different hurricanes but differently at the geographical regions. (b) Both rapid- and slow-recovery are present for Hurricane Ike but only slow recovery is shown for a regional distribution network from Hurricane Sandy.

%Title, abs. Dependence?
%Ling's comments

%-------
%Redo fig. 7band 8: Use 9 cities with 16 or more samples. Also, re-enumerate the cities so that the
%number in Fig. 7 is sequential with the timing of the peaks, and also consistent to those used in Fig. 8

%Spatial model is never tested. How to test it? (a) Can observe from the data whether the failure durations are stationary within a county. (b) May combine nearby counties to obtain more samples for a bigger region. (c) Need to know how the assumption of region-stationarity deviates from the reality (see from the data).

%Sandy: How to obtain N_f, N_r (see Fig. 10 as an example): Use Ike data and synthetic data to test.

%Need complete formulation for spatial temporal

\end{abstract}


%Knowledge learned: Resilience/vulnerability

\section{Introduction}\label{sec:Intro}

Non-stationary modeling and learning have been widely applied to many applications \cite{Alippi}\cite{sch}. This work contributes a new application in an emerging area of smart energy grid. The application is on learning from failure data how distributed power networks respond to external disturbances such as hurricanes. Learned knowledge provides understanding how power networks fail and recover in severe weather. Such understanding is a prerequisite of modernizing our power infrastructure.

Power distribution networks lie at the edge of the energy grid, delivering medium and low voltages to residence and organizations \cite{Kaplan09}. Distribution networks consist of leaf nodes of the energy infrastructure and are thus susceptible to external disturbances. For example, natural disasters cause wide-spread destructions and service disruptions to distribution networks \cite{Doereport}\cite{Albert04}. There were about 16 major hurricanes and severe storms occurred in north America in the past 5 years \cite{Wiki}, each of which disrupted electricity services from 500,000 to several million customers for days \cite{Wiki}.

Existing approaches rely primarily on empirical approaches for large-scale failures of power distribution. For example, empirical studies have been conducted on assessing damages from large-scale power failures (see \cite{Doereport1} and references therein). Monitoring systems have been used in power industry to respond to failures (see \cite{Entergy11} as examples). As hurricanes and severe storms appear to occur frequently and at a large-scale \cite{Wiki}, empirical approaches become inadequate for real time failure assessment in a wide geographical area \cite{Doe_sandy}. Furthermore, recovery from large-scale power failures is even less understood. This is evidenced by how difficult it was for utilities to provide accurate recovery time to customers \cite{Doe_sandy}. Overall, quantifiable approaches are lacking and needed for characterizing how power distribution networks respond to external disturbances. This is important for discovering and mitigating vulnerabilities for enhancing the power infrastructure \cite{Rudnick11}\cite{Hooke07}.

Unique challenges emerge for quantifying how power distribution networks respond to large-scale external disturbances. The first is randomness. External disturbances such as hurricanes exhibit random behaviors. The resulting power failures occur randomly also. The second is dynamic nature of failures and recoveries due to evolution of external disturbances. For example, a hurricane usually has a landfall with a strong force wind, and then gradually dies down when moving in land. Hence, non-stationarity (randomness and dynamics) is an intrinsic characteristic of large-scale failures.

Non-stationary learning is a natural approach for quantifying non-stationary large-scale failure and recovery of power distribution induced by external disturbances. However, an additional challenge for learning is lack of data. This may appear to be a paradox: A large-scale external disturbance such as a hurricane often results in thousands of power failures, which amounts to a lot of data. However, in the space of external disturbances, a hurricane generates only one sample, i.e., a snap-shot of network failures and recoveries from one external disturbance. Hence, data from an individual disturbance is valuable and should be used to enable learning. Note that using real data for studying large-scale power failures and recoveries is not yet a common practice for the power infrastructure. Real data on power failures from external disturbance is rare \cite{Zhu}\cite{Liu_Singh_11}. A recent work shows the strength of combining algorithmic approaches with real data on geo-graphically correlated power failures \cite{Bernstein12}. The focus there is on power transmission rather than distribution.

Incorporating all challenges, a basic research question we intend to answer is how to learn non-stationary behaviors of large-scale failure and recovery for distributed power distribution, using real data from one external disturbance? Combining model-based and data-driven methods is a viable approach for limited samples \cite{bienenstock91}. A model identifies pertinent quantities that determine non-stationary random processes of failure and recovery. We first derive a problem formulation to obtain a model. What remains unknown are model parameters, which can be learned from data. Such a combination of model-based and data-driven approaches directs learning to a small number of functions or parameters, and thus makes effective use of data. In addition, a combination of model-based and data-driven approaches makes learning explanatory: Learned model parameters bear physical meanings on how distributed power distribution responds to an external disruption.

Our formulation focuses on power failures and recoveries induced by exogenous weather. The time scale of such failures and recoveries is considered to be a minute to be consistent to that of a hurricane (see Section \ref{sec:ike} for details). Power failures can also occur in bursts at a small time scale of seconds or less \cite{Amin08}. Such bursty failures are usually due to an internal network structure (see Section \ref{sec:ike}) and not studied in this work. Self-recoveries often occur at the small time scale of sub-seconds \cite{Amin08} whereas recovery by field crews occur in minutes or beyond. Hence our model at the time scale of a minute focuses on weather induced failures and recoveries that can not be repaired through self-healing. Such a model provides understanding how distributed power infrastructure responds to external disturbances.

Our formulation begins with the spatial scale of network nodes and the temporal scale of a minute. As the data from an external disturbance is insufficient to completely specify a detailed temporal-spatial model \cite{Duda}, we aggregate spatial variables into groups. A group can be a city that consists of nodes from a small geo-graphical area. The resulting model thus characterizes an entire non-stationary life-cycle of large-scale failure and recovery in time and at geo-locations. Such a spatial-temporal model is multivariate generalization of $GI(t)/G(t)/ \infty$ queues \cite{Bertsimas97} to include geo-locations. $GI(t)$'s and $G(t)$'s are arrival (failure) processes and departure (recovery) processes for individual geo-graphical area. ``$\infty$'' means that it is possible for recovery to occur immediately after a failure, e.g., less than a minute in this work. Hence, multivariate $GI(t)$'s and $G(t)$'s constitute our model that completely specify non-stationary behaviors of large-scale failure and recovery at a power distribution network.

We consider one simplified characterization of $GI(t)/G(t)/\infty$ queues to the expected values \cite{Bertsimas97}. What to learn then becomes clear: A small number of pertinent parameters of $GI(t)$ and $G(t)$ at different geo-locations, i.e., failure rates and recovery time distributions. We first obtain detailed data on large-scale power failures from a real life example of a natural disaster, Hurricane Ike. Ike caused power failures in the south states of US and affected more than 2 million users in 2008. We devise learning for two scenarios using the real data. The first learns only temporal processes of non-stationary failure and recovery by aggregating over spatial variables of nodes in an entire network. The second learns geo-location based spatial-temporal processes by aggregating nodes in cities. We show the modeling facilitates learning where model parameters can be easily estimated using the failure data. We then apply the model to another data set from Hurricane Sandy. Hurricane Sandy caused wide-spread power failures to more than 8 million people in the northeast of US in 2012. The data set consists of aggregated rather than detailed power failures in one of the impacted areas. Our approach is shown to be applicable to the aggregated data for estimating failure and recovery rates. Our approach also shows what can not be learned using aggregated data.

In summary, the contribution of this work consists of the following: (a) a novel model based on non-stationary random processes and dynamic queues for weather-induced large-scale failure and recovery of power distribution, (b) simple learning approaches for estimating parameters of the non-stationary model, (c) applications of the model and non-stationary learning to real data from two hurricanes at different locations.

The rest of the paper is organized as follows. Section \ref{sec:Background} provides background knowledge and an example of large-scale failures at a power distribution network. Section \ref{sec:StochasticModel} and \ref{sec:FRProc} develops a problem formulation of spatial-temporal non-stationary random processes. Section \ref{sec:ike} describes the real data from Hurricane Ike and learns a geo-temporal model. Section \ref{sec:sandy} studies non-stationary failure and recovery using parts of real data from Hurricane Sandy. Section \ref{sec:discuss} discusses our findings. Section \ref{sec:Conclusion} concludes the paper.

\section{Background and Example}\label{sec:Background}

We now provide examples on the temporal scale, and non-stationarity of failure and recovery.

\subsection{Time Scale of Failure and Recovery}

We first discuss the time scale for modeling weather induced failures and recoveries. A power distribution network consists of components such as substations, feeders, transformers, power circuits, circuit breakers, transmission lines, and meters. An example power distribution system is illustrated in Figure \ref{fig:distnet}, with a commonly used radial topology. Three types of components are shown for illustration: A primary substation, three secondary power sources, and loads. Links correspond to power lines. Assume that either a component or a link can fail during a hurricane. Assume that the substation is used as a primary source during normal operation. The secondary sources, that can be distributed renewable sources, are used for back-up when the primary source fails \cite{perry12}. Then the following scenarios can occur for failure and recovery:

\begin{figure}
  \centering
  \includegraphics[width=0.4\textwidth]{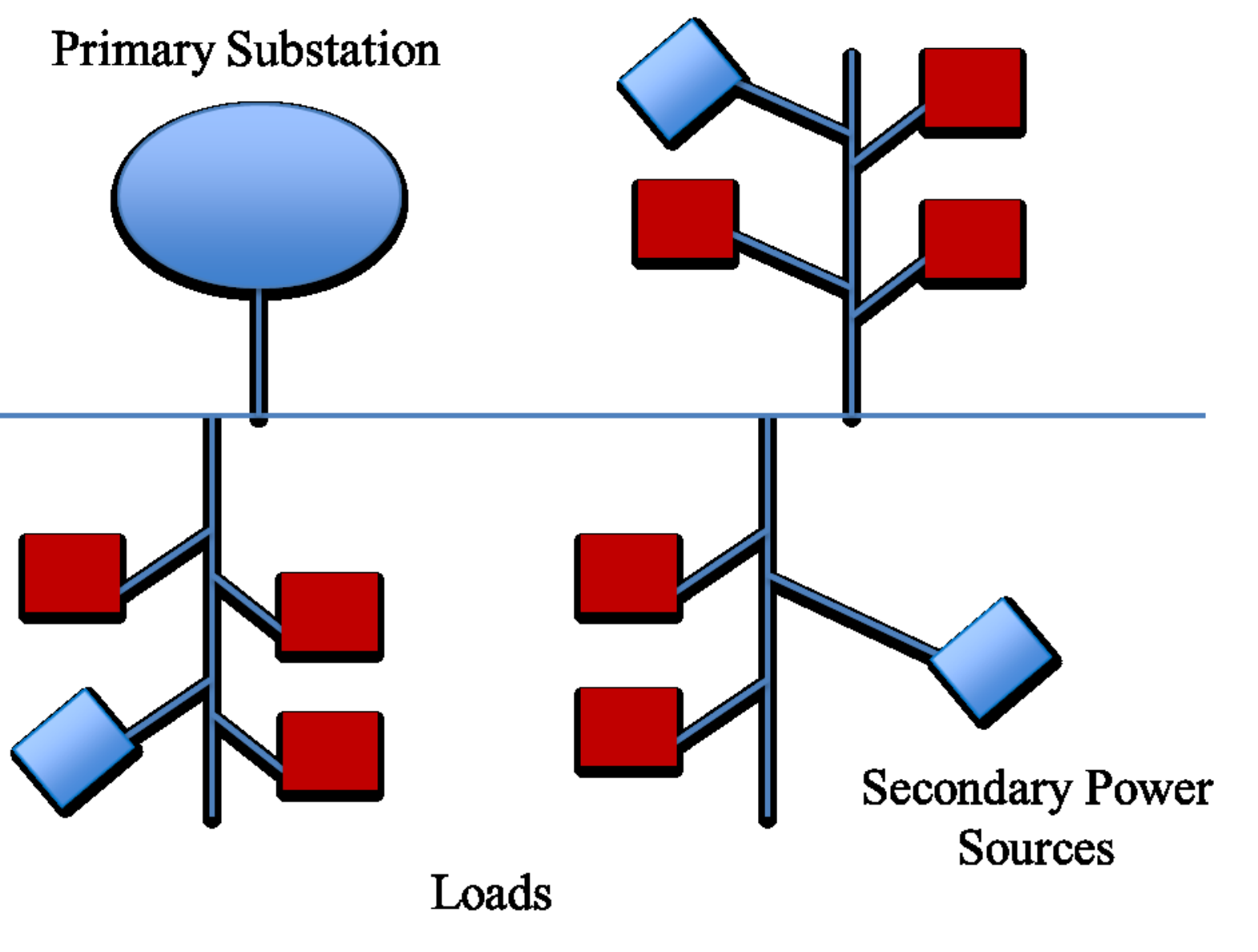}
  \caption{A Section in A Distribution Network.}
  \label{fig:distnet}
\end{figure}

(a) If all the sources fail due to an external disturbance, there is no electricity supply to
any loads. Hence, the loads experience dependent failures that can occur instantaneously. The scenario of dependent failures also applies to other components upstream in a radial topology that cause loss of electricity at nodes downstream. Dependent failures are often experienced by loads within sub-seconds.

(b) If a link that connects a load to the network fails due to an external disturbance, there is no electricity supply to the load. Such link failures can occur independently due to fallen trees or power lines. Thus loads experience independent loss of electricity. As such independent failures are caused by exogenous weather, they are assumed to occur at a time scale of a minute or beyond. Such a time scale can be estimated through how rapidly a hurricane force wind passes a city. Consider a small city of $1,600$ acres as an example. Based on the IEEE standard (IEEE/ASTM SI 10-1997)\cite{ieee}, an approximated ``diameter'' of the city is about $1.6$ miles. Consider the speed of the force wind at $60$ miles per hour. It takes about $1.6$ minutes for the wind to pass the city. This provides a basis of using a minute as a time scale of weather-induced failures.

(c) Recovery depends on the types of failures and recovery schemes. Certain failures can be repaired through self-recovery \cite{Amin08}. For example, if the primary substation fails, the electricity supply to all loads can be recovered when the three secondary sources are in operation. In general, self-recovery and automated reconfiguration built in power distribution usually operate at a time scale of sub-seconds or seconds \cite{Amin08}. However, failures due to external disturbances, e.g., falling trees and power lines, often require manual repair by field crews. Recovery time depends on not only restoration schemes but also environmental constraints, and is thus considered as random in this work. Such manual recovery time is in either minutes or hours or days from failures.

In summary, failures and self-recoveries at a small time-scale of seconds or sub-seconds depend on detailed network structure and self-recovery schemes. Failure and recovery at a larger time scale of a minute and beyond are often due to external disturbances that evolve dynamically and randomly.

\subsection{Example of Non-Stationary Failure and Recovery}

\begin{figure}
\begin{center}
\includegraphics[width=0.4\textwidth]{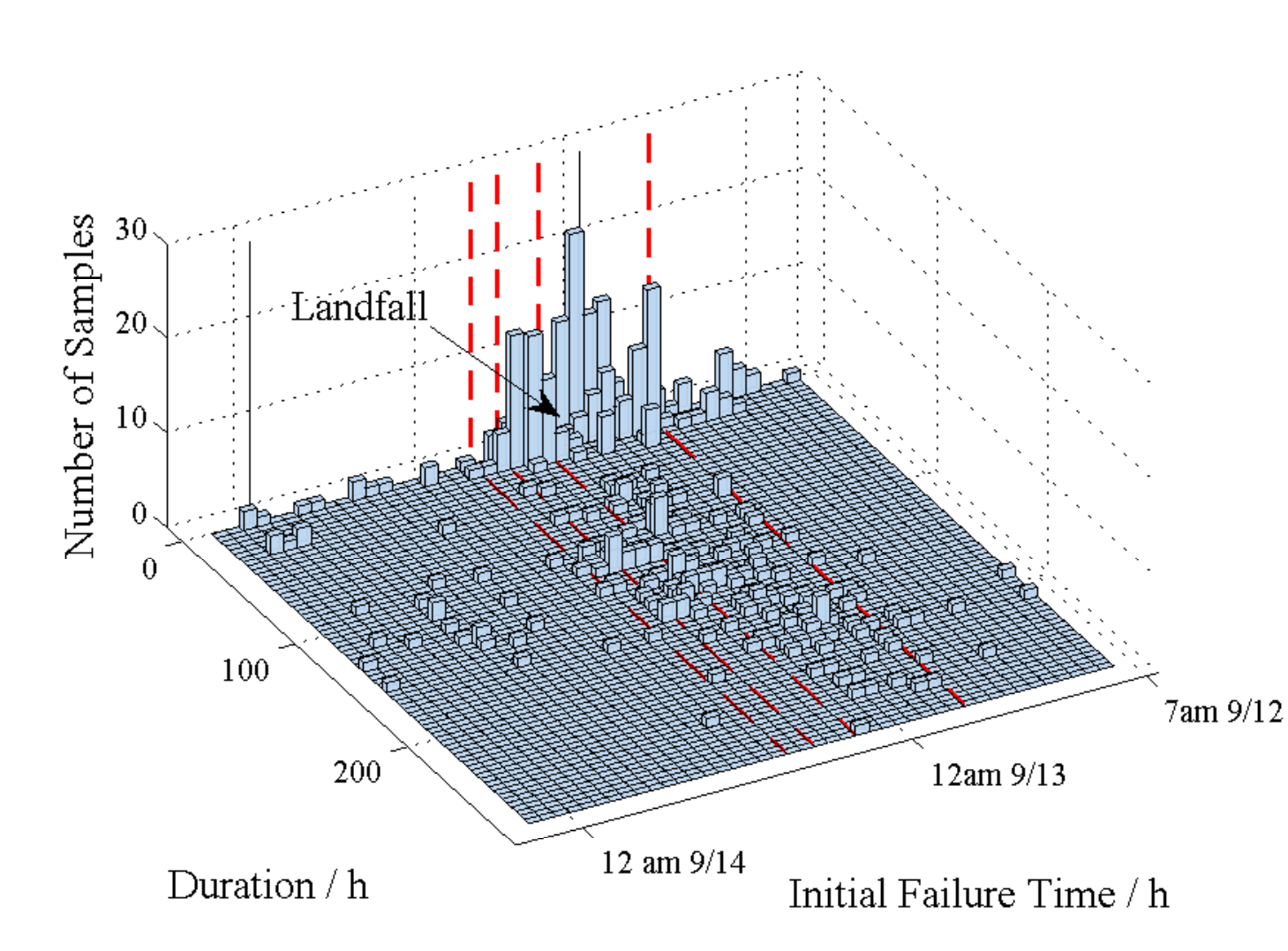}
\caption{Empirical temporal distribution of failure durations in 3D.}
\label{fig:HistPDur}
\end{center}
\end{figure}

\begin{figure}
\begin{center}
\includegraphics[width=0.4\textwidth]{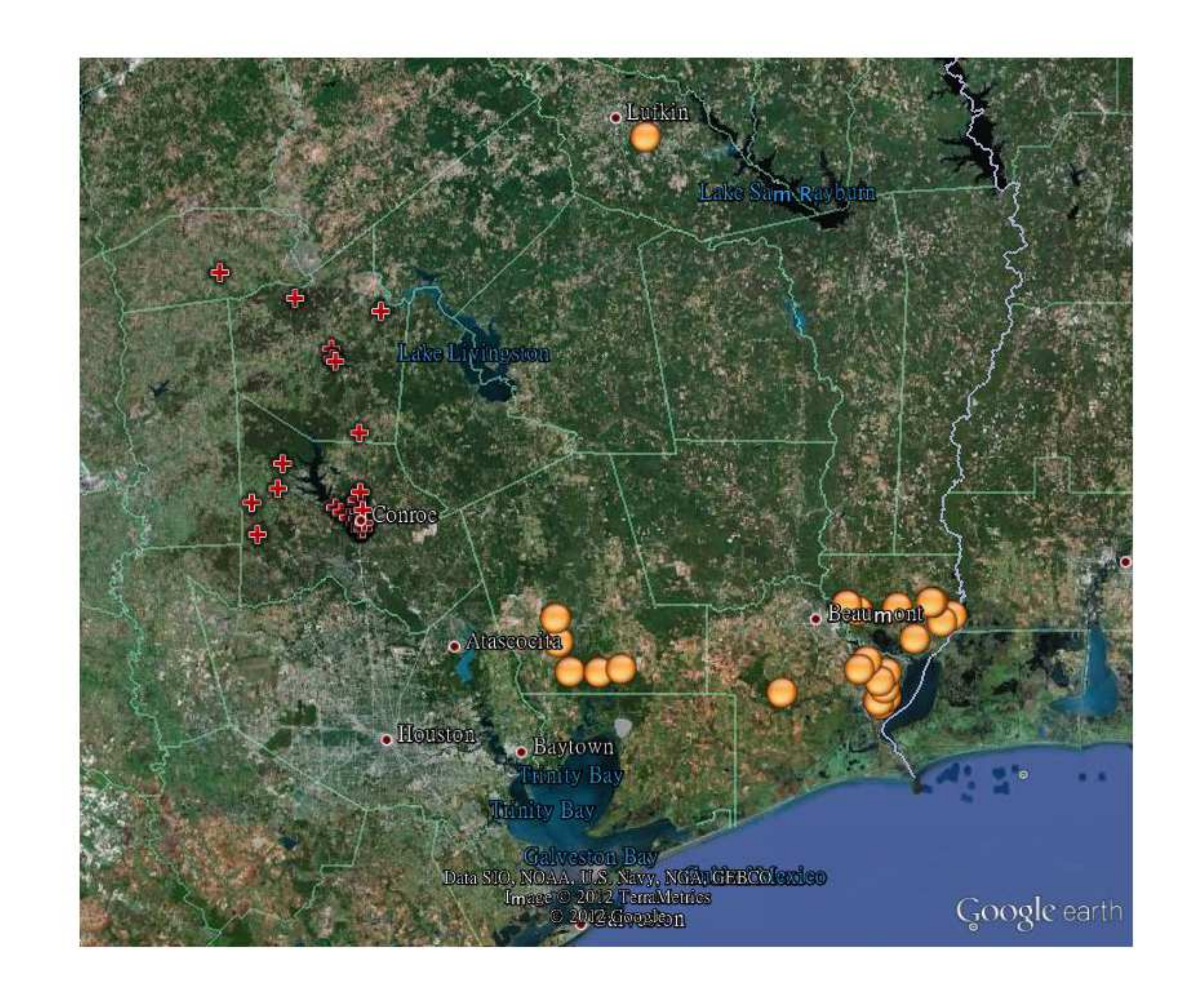}
\caption{Geo-locations of failures occurred in different time durations. Red marker: Failures occurred from 7 p.m to 8 p.m. Sep. 12. Yellow marker: Failures occurred from 5 a.m. to 6 a.m. Sep. 13.}
\label{fig:GeoP}
\end{center}
\end{figure}

To gain intuition on an entire life cycle of failure and recovery of a distribution network, we consider a real-life example of large-scale power failures occurred during Hurricane Ike in 2008. Figure \ref{fig:HistPDur} shows a histogram on failure occurrence time and duration at an operational distribution network before, during and after the hurricane. Each bin has length (failure occurrence time) of $1$ hour\footnote{CDT is used for all plots for Hurricane Ike.} and width (duration) of $4$ hours. The height of each bin represents the number of failures that occur at time $t$ and last for duration $d$. Figure \ref{fig:GeoP} shows geographical distributions of failure occurrences at two different time epochs, where failure occurrence is evidently non-stationary across geographical regions. Hence,

(a)	Failure occurrence is non-stationary, i.e., random and time-varying;

(b)	Recovery time is non-stationary, i.e., obeys different probability distributions for failures occurred at different time;

(c)	Failure occurrence and recovery time are also non-stationary spatially, i.e., exhibit different distributions for different geo-locations.

Hence, samples on failure occurrence time and duration are not identically distributed but exhibit geo-temporal non-stationarity.

\subsection{Non-Stationary Learning}

Non-stationary random processes have been studied in the context of drifting concepts (see \cite{kuh}\cite{widmer}\cite{gama}\cite{Elwell} and references therein). Samples for learning are dynamically drawn from a non-stationary environment. An issue arises on the sample size, i.e., whether data is sufficient for characterizing underlying drifts of distributions.

The problem of learning non-stationary processes in this work exhibits unique challenges in terms of sample size. For simplicity, batch data is assumed to be collected for learning an entire non-stationary life cycle of failure and recovery processes off-line. A challenge here is that there is only one snapshot of a distribution network in space and time from one external disturbance. The number of data sets is often small, i.e., from a few severe storms. Therefore, combining model-based and data-driven approaches becomes important, where data can be used to learn a small number of model-parameters from one external disturbance at a time \cite{bienenstock91}. In addition, combining model-based and data-driven approaches for learning is required by the problem: Learned model parameters need to exhibit physical meaning for generic network behaviors upon external disturbances.

\section{Stochastic Model}\label{sec:StochasticModel}

We now formulate large-scale failure and recovery based on non-stationary random processes. We begin with the detailed information on nodal statuses in a distribution system. We then aggregate the spatial variables of nodes to obtain temporal evolution of failure and recovery across geo-graphical areas.

\subsection{Failure and Recovery Probability}

A geo-temporal random process provides a theoretical basis for modeling large-scale failures. The temporal variable is time $t$ that is assumed to be continuous at the scale of a minute. The spatial variable can be either geo- or network-location of a node. For simplicity, this work considers geo-location as a spatial variable to focus on location-based failures induced by severe weather. Nodes can be components in a distribution system such as substations, feeders, hubs, transformers, transmission lines, and distributed energy sources. A shorthand notation $i$ is used to specify the index of node $i$ located at $z_i$. $i\in S=\{1,2,...,n\}$ for a power distribution network with $n$ nodes. An underlying network topology is assumed to be radial so that cascading failures occurred in mesh networks are not considered.

Let $X_i(z_i, t)$ be the status of the $i$-th node at time $t>0$ for $1 \le i \le n$. We assume for simplicity that nodes only exhibit two states: $X_i(z_i,t)=1$ if the $i$-th node is in a failure mode, i.e., without power supply. $X_i(z_i,t)=0$ if the node is in normal operation. Failures caused by external disturbances exhibit randomness. Whether and when a node fails is random. Whether and when a failed node recovers is also random. Hence, random processes can be used to characterize failure and recovery for all nodes in a network.

Given time $t>0$, $P\{X_i(z_i, t+\tau)=1\}$ characterizes the probability that node $i$ is failed in the near future $t+\tau$, where $\tau>0$ is a small time increment. Assume a node changes state, i.e., from failure to normal and vice versa. Then for the $i$th node, $1 \le i \le n$, the probability that node $i$ stays in failure mode in $[t,t+\tau]$ is,
\begin{equation}
\begin{split}
&P\{X_i(z_i,t+\tau)=1\}-P\{ X_i(z_i, t)=1\} \\
=&P\{X_i(z_i,t+\tau)=1,X_i(z_i,t)=0\} \\
&-P\{X_i(z_i,t+\tau)=0,X_i(z_i,t)=1\}. \\
\end{split}
\label{eq:StateTrans1}
\end{equation}

Equation \ref{eq:StateTrans1} assumes Markov temporal dependence, and can be applied to $n$ nodes in a distribution network.  The $n$ equations together form a geo-temporal model of a network. Note that statistically dependent failures at the small time scale less than a minute are not considered here, as such failures are often caused by an internal network structure rather than exogenous weather. Spatial dependence is embedded in the model but will be studied explicitly in subsequent work.

\begin{figure}
\begin{center}
\includegraphics[width=0.45\textwidth]{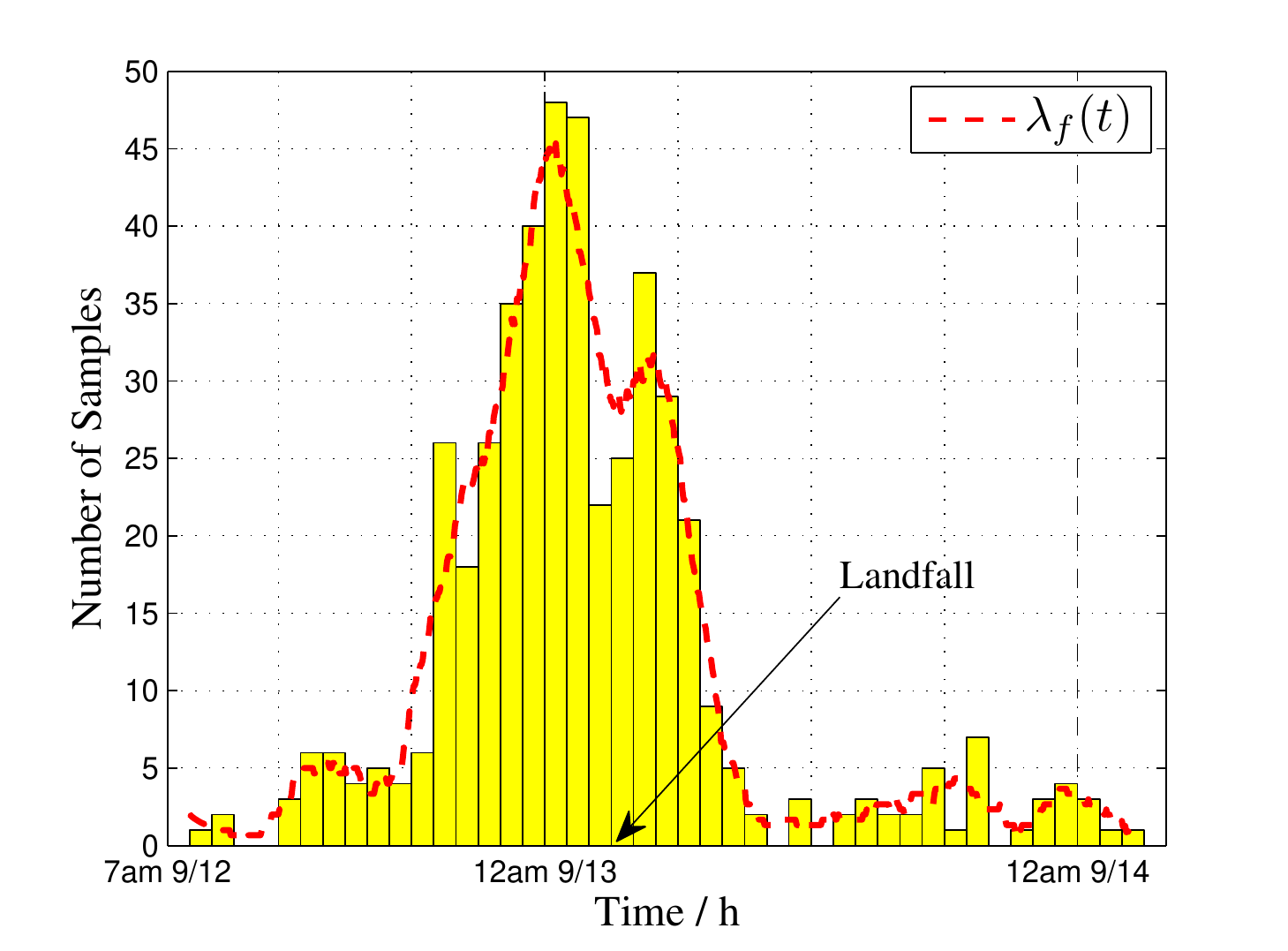}
\caption{Histogram of failure occurrence time and the failure rate $\lambda_f(t)$ during Hurricane Ike.}
\label{fig:histfr}
\end{center}
\end{figure}

\subsection{Aggregated Geo-Temporal Process}

When large-scale failures are caused by one external disturbance, information available is from one ``snapshot'' of temporal spatial network statuses, and thus insufficient for specifying a complete temporal-spatial model at the node level. Hence, nodes are aggregated over a geographical region ($Z$), resulting in

\begin{equation}
\begin{split}
&\sum_{i; z_i \in \mathbb{Z}} P\{X_i(z_i,t+\tau)=1\}-\sum_{i; z_i \in \mathbb{Z}} P\{X_i(z_i,t)=1\} \\
=&\sum_{i; z_i \in \mathbb{Z}} P\{X_i(z_i,t+\tau)=1,X_i(z_i,t)=0\} \\
&-\sum_{i; z_i \in \mathbb{Z}} P\{X_i(z_i,t+\tau)=0,X_i(z_i,t)=1\}. \\
\end{split}
\label{eq:StateTrans2}
\end{equation}

Here $P\{X_i(z_i,t)=1\}=E\{I[X_i(z_i,t)=1]\}$, where $I()$ is an indicator function. $I(A)=1$ if event A occurs, and $I(A)=0$ otherwise. We can define a geo-temporal process as follows.

\textbf{Definition}: $\{N(t,\mathbb{Z}) \in \mathds{N},t>0\}$ is a geo-temporal process where the spatial variables ($i$'s) are aggregated for all nodes $z_i$ in a predefined region $ \mathbb{Z}$. $N(t,\mathbb{Z})$ is the number of nodes in failure state at time $t$ located in $\mathbb{Z}$,
\begin{equation}
N(t,\mathbb{Z})=\sum_{i; z_i \in \mathbb{Z}} I[X_i(z_i,t)=1].
\label{eq:frpDef}
\end{equation}

Combining Equations \ref{eq:StateTrans2} and \ref{eq:frpDef}, we have,
\begin{equation}
\begin{split}
E\{\Delta N(t,\mathbb{Z})\} =& \sum_{i;z_i \in \mathbb{Z}} P\{X_i(z_i,t+\tau)=1\} \\
& \qquad - \sum_{i; z_i \in \mathbb{Z}} P\{X_i(z_i, t)=1\},
\end{split}
\label{eq:frpExp}
\end{equation}
where $\Delta N(t,\mathbb{Z})=N(t+\tau,\mathbb{Z})-N(t,\mathbb{Z})$ is an increment of the number of failed nodes in a certain region. $\Delta N(t,\mathbb{Z})$ is the result of either newly-failed or newly-recovered nodes. Hence, we define a failure process and a recovery process respectively.

\textbf{Definition}: Failure process $\{N_f(t,\mathbb{Z}) \in \mathds{N},t\ge 0\}$ is the number of failures occurred up to time $t$. Recovery process $\{N_r(t,\mathbb{Z}) \in \mathds{N},t\ge 0\}$ is the number of recoveries occurred up to time $t$.

Assume $\tau>0$ is sufficiently small so that failure or recovery occurs at most once to a node during $(t, t+\tau)$. The increments on a failure process and a recovery process satisfy respectively,
\begin{equation}
\begin{split}
E\{\Delta N_f(t, \mathbb{Z})\} &= \sum_{i;z_i \in\mathbb{Z}} P\{X_i(z_i,t+\tau)=1, X_i(z_i,t)=0\}, \\
E\{\Delta N_r(t, \mathbb{Z})\} &= \sum_{i;z_i \in\mathbb{Z}} P\{X_i(z_i,t+\tau)=0, X_i(z_i,t)=1\}, \\
\end{split}
\end{equation}
where $\Delta N_f(t,\mathbb{Z})=N_f(t+\tau,\mathbb{Z})\} - N_f(t,\mathbb{Z})$. Similarly, for a sufficiently small $\tau > 0$, it can be assumed that at most one recovery occurs during $(t,t+\tau)$. Hence, Equation \ref{eq:StateTrans2} is simplified as,
\begin{equation}
E\{\Delta N(t, \mathbb{Z})\} = E\{\Delta N_f(t, \mathbb{Z})\}- E\{\Delta N_r(t, \mathbb{Z})\}.
\label{eq:StateTrans3}
\end{equation}
Furthermore, we assume at time $t_0=0$, $N(t, \mathbb{Z})=0$, $N_f(t, \mathbb{Z})=0$, and $N_r(t, \mathbb{Z})=0$. Aggregating increments in Equation \ref{eq:StateTrans3} from $0$ to $t$, we have,
\begin{equation}
E\{N(t, \mathbb{Z})\} = E\{N_f(t, \mathbb{Z})\}- E\{N_r(t, \mathbb{Z})\}.
\label{eq:FRE}
\end{equation}

Hence, the expected number of nodes in the failure state equals to the difference between the expected failures and the expected recoveries. We now group a distribution network of $n$ nodes into $m$ geographical regions $\mathbb{Z}_j$, $1 \le j \le m$, based on their geo-locations. A city, e.g., a subdivision, is an example of a geo-graphical region widely-used by utilities. Then the failure-recovery process for the entire distribution network $N(t)$ is defined as,
\begin{equation}
N(t) = [N(t, \mathbb{Z}_1), N(t, \mathbb{Z}_2), ..., N(t, \mathbb{Z}_m)]^{\mathrm{T}},
\end{equation}
where $N(t, \mathbb{Z}_j)$ characterizes how local power distribution in region $\mathbb{Z}_j$ responds to an external disturbance.

\section{Non-Stationary Failure and Recovery}\label{sec:FRProc}

We now derive non-stationary characteristics on failure and recovery. Our derivation reveals pertinent quantities that completely model the behaviors of large-scale power failures and recoveries in expected values. This is pertinent to learning a small number of parameters in Section \ref{sec:ike}.

\subsection{Failure Process}

A failure process can be characterized to the first moment by failure rate functions. Let $\lambda_f(t)=[\lambda_f(t, \mathbb{Z}_1)$, $\lambda_f(t, \mathbb{Z}_2)$, $...$, $\lambda_f(t, \mathbb{Z}_m)]^{\mathrm{T}}$ be a vector that consists of the rate function of a failure process, where $\lambda_f(t, \mathbb{Z}_j)$ is the expected number of new failures per unit time at epoch $t$ and region $\mathbb{Z}_j$, $j=1,2,...,m$,
\begin{equation}
\lambda_f(t, \mathbb{Z}_j)=\lim_{\tau \rightarrow 0} \frac {1} {\tau} E\{N_f(t+\tau, \mathbb{Z}_j)-N_f(t, \mathbb{Z}_j)\}.
\label{eq:FRateDef}
\end{equation}
The larger $\lambda_f(t, \mathbb{Z}_j)$ is, the faster failures occur in $\mathbb{Z}_j$ at time $t$. $\lambda_f(t,\mathbb{Z}_j)$ is referred to as the rate function of the failure process $N_f(t,\mathbb{Z}_j)$. Hence, failure rate quantifies the intensity of failure occurrence. An non-stationary failure process has a time-varying intensity function $\lambda_f(t,\mathbb{Z}_j)$ across geo-locations. Assuming a failure process begins at $t=0$, we have $ E\{N_f(t)\} = [E\{N_f(t, \mathbb{Z}_1)\}, ..., E\{N_f(t, \mathbb{Z}_m) \}]^{\mathrm{T}}$, where
\begin{equation}
E\{N_f(t, \mathbb{Z}_j)\}=\int_0^t \lambda_f(v, \mathbb{Z}_j)dv,
\label{eq:fp}
\end{equation}
for $ 1 \le j \le m$.

\subsection{Recovery Process}

A recovery process can be characterized by recovery rate function $\lambda_r(t)$, where
$\lambda_r(t)=[\lambda_r(t, \mathbb{Z}_1),$ $\lambda_r(t, \mathbb{Z}_2),$ $...,$ $\lambda_r(t, \mathbb{Z}_m)]^{\mathrm{T}}$. $\lambda_r(t, \mathbb{Z}_j)$ is the expected number of new recoveries per unit time at epoch $t$ and region $\mathbb{Z}_j$,
\begin{equation}
\lambda_r(t, \mathbb{Z}_j)=\lim_{\tau \rightarrow 0} \frac {1}{\tau} E\{N_r(t+\tau, \mathbb{Z}_j) - N_r(t, \mathbb{Z}_j)\}.
\label{eq:RRateDef}
\end{equation}
An non-stationary recovery process $N_f(t, \mathbb{Z}_j)$ has a time-varying rate function. Assuming the temporal failure process begins at $t=0$, we have for $1 \le j \le m$,
\begin{equation}
E\{N_r(t,\mathbb{Z}_j)\}=\int_0^t \lambda_r(v,\mathbb{Z}_j)dv.
\label{eq:rp}
\end{equation}

The recovery rate characterizes how rapidly recovery occurs, which is measured by failure duration $D$. For an non-stationary recovery process, a failure duration depends on when and where a failure occurs as illustrated in Figure \ref{fig:HistPDur}. Such non-stationarity of recovery is characterized by $g(d|t, \mathbb{Z}_j)$ which is a conditional probability density function of failure duration $D=d$ given failure time $T=t$ at region $\mathbb{Z}_j$. For a given threshold $d_0>0$, the conditional probability that a duration is bounded by $d_0$ for failures occurred at time $t$ is

\begin{equation}
P\{D < d_0|t, \mathbb{Z}_j\}=\int_0^{d_0} g(v|t, \mathbb{Z}_j) dv.
\label{eq:pd}
\end{equation}

When $d_0$ is sufficiently small, this probability characterizes rapid recovery that occurs shortly after failures. For a given $d_0$, the larger $P\{D < d_0|t, \mathbb{Z}_j\}$ is, the more rapid recovery dominates a recovery process. Given desired value of probability $P\{D < d_0|t, \mathbb{Z}_j\}$, the smaller $d_0$ is, the more dominating the rapid recovery is.

Rapid recovery is referred to as infant recovery. This terminology is borrowed from infant mortality in survivability analysis \cite{Hosmer08}. Infant recovery is a desirable characteristic of the smart grid. In contrast, slow recovery is referred to as aging recovery in analogous to aging mortality \cite{Kalbfleisch02}. Infant and aging recovery can be formally defined as follows.

\textbf{Definition}: Let $d_0>0$ be a threshold value. If a node remains in failure for a duration less than $d_0$; a recovery is an infant recovery. Otherwise, the recovery is aging recovery. Infant recovery is characterized by $P\{D < d_0|t, \mathbb{Z}_j\}$. Aging recovery is characterized by $P\{D > d_0|t, \mathbb{Z}_j\}$.

\subsection{Joint Failure-Recovery Process}

A joint failure-recovery process characterizes an entire life cycle of a failure-recovery process (FRP), and represents the total number of nodes $N(t,\mathbb{Z})$ in failure state at time $t$ in region $\mathbb{Z} $(Equation \ref{eq:frpDef}). The expected number of nodes in failure can be expressed in rate functions,
\begin{equation}
E\{N(t,\mathbb{Z}_j)\}=\int_{0}^t [\lambda_f(v,\mathbb{Z}_j)-\lambda_r(v,\mathbb{Z}_j)]dv.
\label{eq:frp}
\end{equation}

Failure-and-recovery process can be viewed as a birth-death process. However, commonly used birth-death processes have a stationary distribution of failure duration and assume independence between failure occurrence $t$ and failure duration $d$ \cite{Ross10}. Here, these two assumptions do not hold. This implies that failures occurred at different time can last different duration. For example, under strong and sustained hurricane wind, failures that do not happen in day-to-day operation can occur due to falling debris and power lines. We shall further elaborate this through the real-life examples in Sections \ref{sec:ike} and \ref{sec:sandy}.

A recovery process is related to a failure process through a probability density function of failure durations.

\indent

\textbf{Theorem}\label{thm:rp}
\textit{Let $\{N_f(t,\mathbb{Z}_j)\}$ be an independent increment (failure) process with a rate function $\lambda_f(t,\mathbb{Z}_j)$, $1 \le j \le m$. Let $D(t)$ be the duration of a failur occurred at time $t$ and region $\mathbb{Z}_j$. $D(t)$ has a conditional probability density function $g(d|t,\mathbb{Z}_j)$, where $d \ge 0$, $t \ge 0$.
Then recovery rate $\lambda_r(t,\mathbb{Z}_j)$ satisfies}
\begin{equation}
\lambda_r(t,\mathbb{Z}_j)=\int_0^{t} g(t-s|s,\mathbb{Z}_j) \lambda_f(s,\mathbb{Z}_j) ds,
\label{eq:RRate}
\end{equation}
\textit{where $1 \le j \le m$, $d=t-s$ with $s$ and $t$ being the failure time and recovery time respectively.}

\indent

The theorem is a corollary of the Transient Little's Theorem \cite{Bertsimas97}. Intuitively, $g(t-s|s,\mathbb{Z}_j) ds$ can be viewed as the probability that a failure occurred at time $s$ and region $\mathbb{Z}_j$ lasts $t-s$ duration. $g(t-s|s,\mathbb{Z}_j) ds \lambda_f(s,\mathbb{Z}_j)$ is the average number of failures per unit time recover after $t-s$ duration, i.e., the recovery rate by definition. Aggregating over all failures occurred prior to time $t$ results in Equation \ref{eq:RRate}. The detailed proof is given in \cite{Wei12}.

\subsection{What to Learn}

What to learn now becomes apparent. Failure rate functions and probability density functions of recovery time completely specify our model to the first moment, i.e.,

\begin{itemize}

\item $\lambda_f(t, |\mathbb{Z}_j)$, for $ 1 \le j \le m$,

\item $g(t-s|s, \mathbb{Z}_j)$, for $1 \le j \le m$.

\end{itemize}

In general, the forms and the parameters of these two functions are unknown, and need to be learned from real data. The learned functions and the parameters can then be used to estimate the empirical processes. The empirical processes are the sample means $\hat N(t, \mathbb{Z}_j)$, $\hat N_f(t, \mathbb{Z}_j)$, and $\hat N_r(t, \mathbb{Z}_j)$ that estimate the true expectations $E\{N(t,\mathbb{Z}_j)\}$, $E\{N_f(t,\mathbb{Z}_j)\}$, and $E\{N_r(t,\mathbb{Z}_j)\}$, respectively.

\section{Hurricane Ike}\label{sec:ike}

We first apply learning to a real-life example of large-scale utility-service disruptions caused by a hurricane.

\subsection{Data From Hurricane Ike}

Hurricane Ike was one of the strongest hurricanes occurred in 2008. Ike caused large scale power failures, resulting in more than 2 million customers without electricity, and marked as the second costliest Atlantic hurricane of all time \cite{Blake11}\cite{fema08}.

Reported by National Hurricane Center \cite{nhc_ike_report}, the storm started to cause power failures across the onshore areas in Louisiana and Texas on September 12, 2008 prior to the landfall. Ike then made a landfall at Galveston, Texas on 2:10 a.m. (CDT), September 13, 2008, causing strong winds, flooding, and heavy rains across Texas. The hurricane weakened to a tropical storm at 1:00 p.m. September 13 and passed Texas by 2:00 a.m. September 14.

A major utility provider collected data on power failures from more than ten cities. The failures include failed circuits, fallen poles and power lines, and non-operational substations. The raw data set has of 5152 samples. Each sample consists of the failure occurrence time ($t_i$) and duration ($d_i$) of a component ($i$) in a distribution network from September $12$ through $14$, 2008. The accuracy for time $t$ is a minute.

\subsection{Data Processing}

The data set contains bursts of failures that occurred within a minute. As a minute is the smallest time scale for each sample, the bursts are considered as dependent failures. Dependent failures are grouped as one failed entity ($i$), with a unique failure occurrence time $t_i$ and duration $d_i$. After such preprocessing, the resulting data set has 465 failed entities. Two outliers with negative failure duration are further removed. The remaining 463 failed entities from 7 am September 12 to 4 am September 14 are referred to as nodes. $D=\{t_i, d_i\}_{i=1}^{463}$ is the data set we use for learning.

Spatial variables $\{\mathbb{Z}_i\}$'s can be either chosen a priori or through learning from data. In this work, we choose $\{\mathbb{Z}_i\}$'s to be small cities to include a natural living environment of customers and this method is widely-used by utility providers. There are $13$ cities in the data set as illustrated in Figure \ref{fig:GeoCity}.

%The number of samples for each city is given in Table \ref{tab:histcity}.
%\begin{table}
%  \centering
%  \caption{Number of samples in 13 cities.}
%  \label{tab:histcity}
%  \begin{tabular}{c|c|c|c|c|c|c|c}
%      \hline
%      City & $\mathbb{Z}_1$ & $\mathbb{Z}_2$ & $\mathbb{Z}_3$ & $\mathbb{Z}_4$ & $\mathbb{Z}_5$ & $\mathbb{Z}_6$ & $\mathbb{Z}_7$ \\
%      \hline
%      Samples & 78 & 8 & 61 & 28 & 94 & 101 & 10 \\
%     \hline
%      City & $\mathbb{Z}_8$ & $\mathbb{Z}_9$ & $\mathbb{Z}_{10}$ & $\mathbb{Z}_{11}$ & $\mathbb{Z}_{12}$ & $\mathbb{Z}_{13}$ & \\
%      \hline
%      Samples & 27 & 16 & 6 & 11 & 7 & 16 & \\
%      \hline
%  \end{tabular}
% \end{table}

\subsection{Temporal Failure Process}

We first study the temporal non-stationarity of the failure-and-recovery process. Spatial variables are aggregated across the entire network. This is equivalent to reducing multiple geo-graphical areas to one entire impact-region from the hurricane. Then the geo-temporal failure-recovery process reduces to a temporal process. For notational simplicity, spatial variables are omitted for temporal processes.

The empirical rate function is estimated using a simple algorithm based on moving average \cite{Trees71}: $\hat{\lambda}_f(t)=\frac{\hat N_f(t+\tau)-\hat N_f(t-\tau)}{2\tau}$, where $\tau$ is chosen to be $5$ hours. The resulting rate function is overlaid with the samples on the number of failures $\hat N_f(t)$ in Figure \ref{fig:histfr}, where each bin is of duration 1 hour.

The learned failure rate function shows a time-varying rate of new failure occurrence:

(a) Prior to 7 p.m. September 12, the rate was low, i.e., fewer than 5 new failures occurred per hour. Hence $ 5$ per hour is considered as the failure rate in day-to-day operation.

(b) At 7 p.m. September 12, the rate increased sharply first to 25 new failures per hour. In the next 6 hours, the rate reached the peak value of nearly 50 new occurrences per hour. This is consistent to the weather report \cite{nhc_ike_report} that the strong wind about 145 mph and flooding impacted the onshore areas prior to the landfall. The time of the peak coincides with the landfall at 2:10 a.m 9/13 CDT.

(c) After staying at the high level for about 12 hours (from 7 p.m. September 12 to 7 a.m. September 13), the rate decreased rapidly back to a low level of less than 5 new failures per hours.

\begin{figure}
\begin{center}
\includegraphics[width=0.45\textwidth]{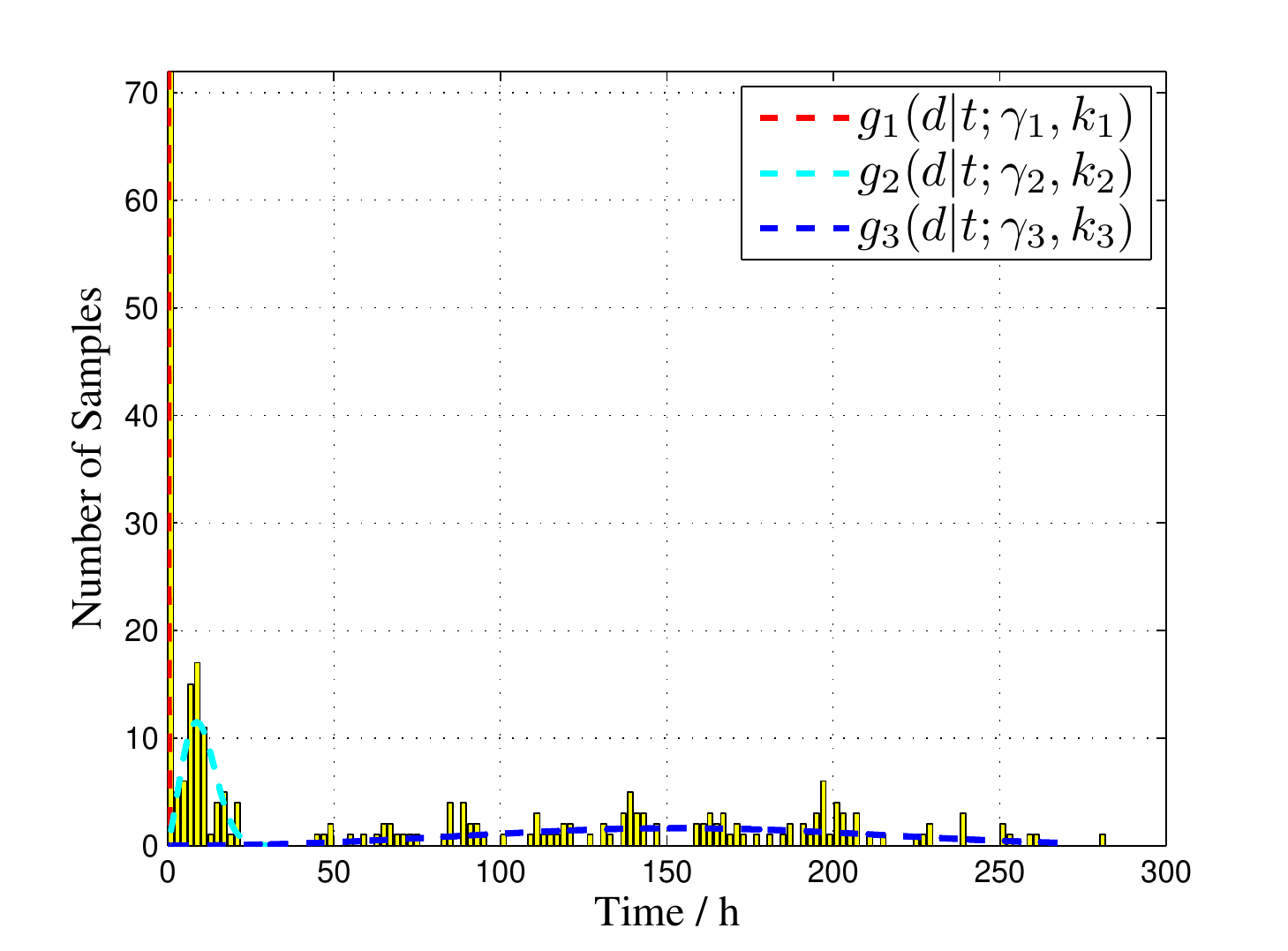}
\caption{Empirical distribution of failure duration for failures occurred during the landfall.}
\label{fig:PDWbl}
\end{center}
\end{figure}

\subsection{Temporal Recovery Process}

We now learn the empirical recovery process characterized by $g(d|t)$, the conditional probability density function of failure duration given failure occurrence time $t$. As the spatial aggregation removes the geo-location variables, $g(d|t)$ is the conditional density function of failure duration of an entire network.

We use the $463$ samples on the failure durations and occurrences in our data set. These samples result in a joint empirical distribution $\hat g(d,t)$ in Figure \ref{fig:HistPDur}. The height of each bin located at $(t,d)$ represents the number of failures that occur at time $t$ and last for duration $d$. Figure \ref{fig:HistPDur} shows non-stationarity of failure durations. For example, a large number (217) of failures occurred between 7 p.m. September 12 and 8 a.m. September 13 lasted for more than a day. This indicates that many failures occurred during the surge of the hurricane were difficult to recover. Hence, a non-stationary distribution for $g(d|t)$ is an appropriate assumption.

\begin{figure}
\begin{center}
\includegraphics[width=0.45\textwidth]{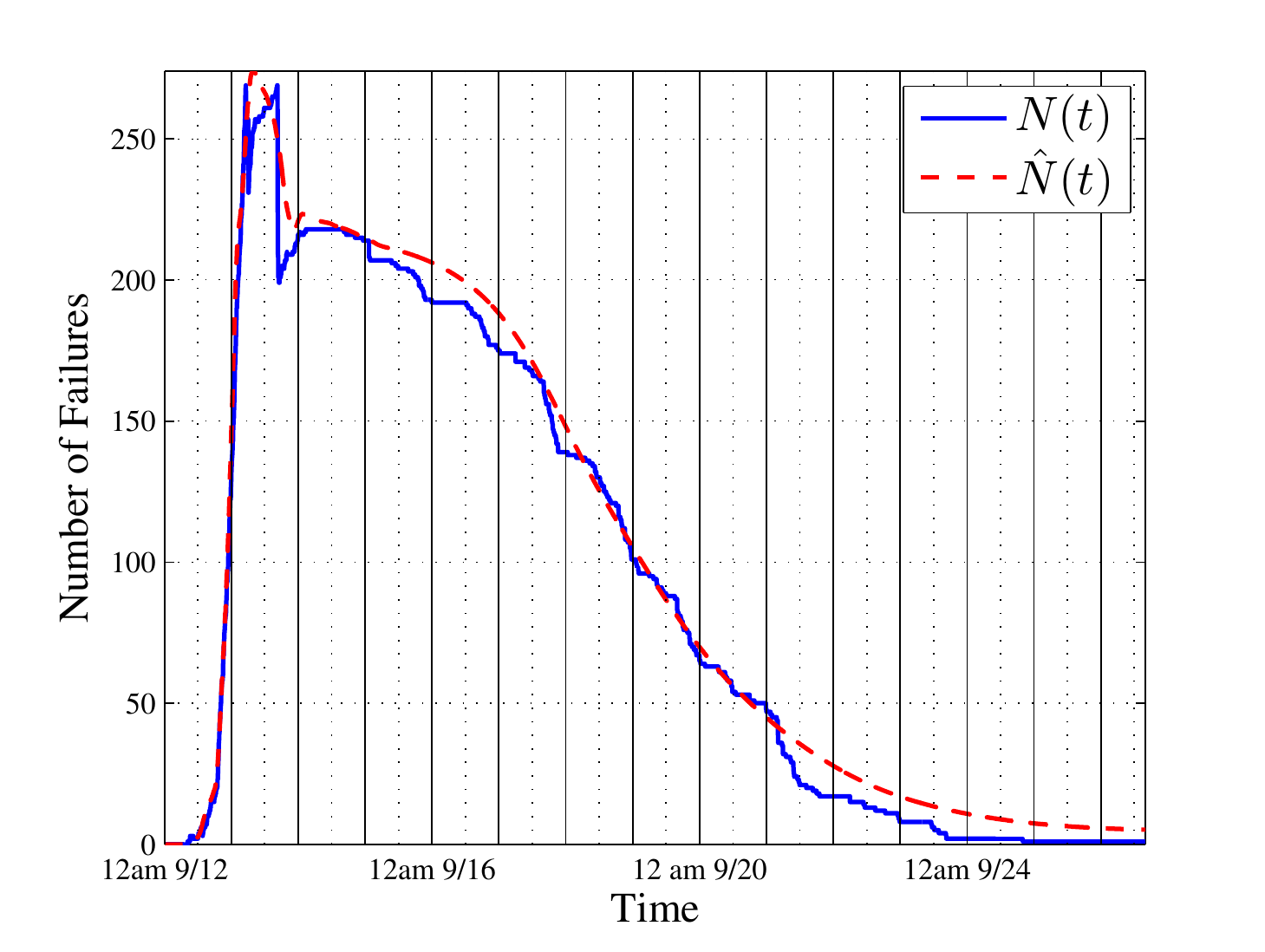}
\caption{Comparison between the joint failure-recovery process $N(t)$ from the data set and the reconstructed process $\hat N(t)$ using learned parameters.}
\label{fig:FRP}
\end{center}
\end{figure}

Given failure occurrence time $t$, we observe that the distribution of duration is a combination of two components: Infant and aging recoveries. We thus select a mixture model for the probability density function $g(d|t)$ where $d >0$,
\begin{equation}
g(d|t)=\sum_{j=1}^{l(t)}{\rho_j (t)g_{j}(d|t)},
\label{eq:mix}
\end{equation}
where $l(t)$ is the number of mixtures at time $t$, $\rho_j(t)$ ($1 \le j \le l$) is a weighting factor for the $j$th mixture function $g_j(d|t)$, and $\sum \rho_j(t)=1$. Weighting factor $\rho_j(t)$ signifies the importance of the $j$th component $g_j(d|t)$. For a non-stationary recovery process, these parameters vary with failure time $t$.

A mixture model is chosen since its parameters exhibit interpretable physical meaning \cite{Chatzis2012}\cite{Fan2012}\cite{Duda}. A parametric family of Weibull mixtures is particularly appealing as the parameters correspond to infant and aging recovery directly. Weibull distributions have been widely used in survival analysis \cite{Kalbfleisch02}\cite{Hosmer08} and reliability theory \cite{Ross10}, but not in characterizing recovery from large-scale external disturbances. Specifically, a Weibull distribution is
\begin{equation}
w(d|t;\gamma(t),k(t))=\frac{k(t)}{\gamma(t)} (\frac{d}{\gamma(t)})^{k(t)-1} e^ {- (\frac {d} {\gamma(t)} )^{k(t)}},
\end{equation}
where $d>0$, $k(t)$ and $\gamma(t)$ are the shape and scale parameters respectively. Hence, $j$th component in Equation \ref{eq:mix} is $g_j(d|t)=w(d|t;\gamma_j(t),k_j(t))$.

Shape and scale parameters, $k(t)$ and $\gamma(t)$, are pertinent for characterizing the type of recovery. The smaller $k(t)$ and $\gamma(t)$ are, the faster the decay of $g(d|t)$, the shorter the failure duration and thus the faster the recovery. Hence, $k(t) < 1$ and moderate $\gamma(t)$ (e.g., $\gamma(t) \sim 10 h$ or smaller) correspond to infant recovery. $k(t) > 1$ and large $\gamma(t)$ (e.g., $\gamma(t) \sim 100 h$) correspond to aging recovery.

For simplicity, we use a piecewise homogeneous function to approximate $g(d|t)$. The failure time $t$ is divided into $5$ intervals shown in Figure \ref{fig:HistPDur}. Within interval $\psi_i$ for $1 \le i \le 5$, $g(d|t\in \psi_i)=g_i(d)$ is assumed to be stationary that does not vary with failure time $t$. For different intervals, $g(d|t \in \psi_i)$'s have different parameters for non-stationarity,
\begin{equation}
g(d|t \in \psi_i)=\sum_{j=1}^{l_i} \rho_{i,j} g_{i,j}(d;\gamma_{i,j},k_{i,j}).
\label{eq:wblpdf}
\end{equation}

\begin{figure}
\begin{center}
\includegraphics[width=0.45\textwidth]{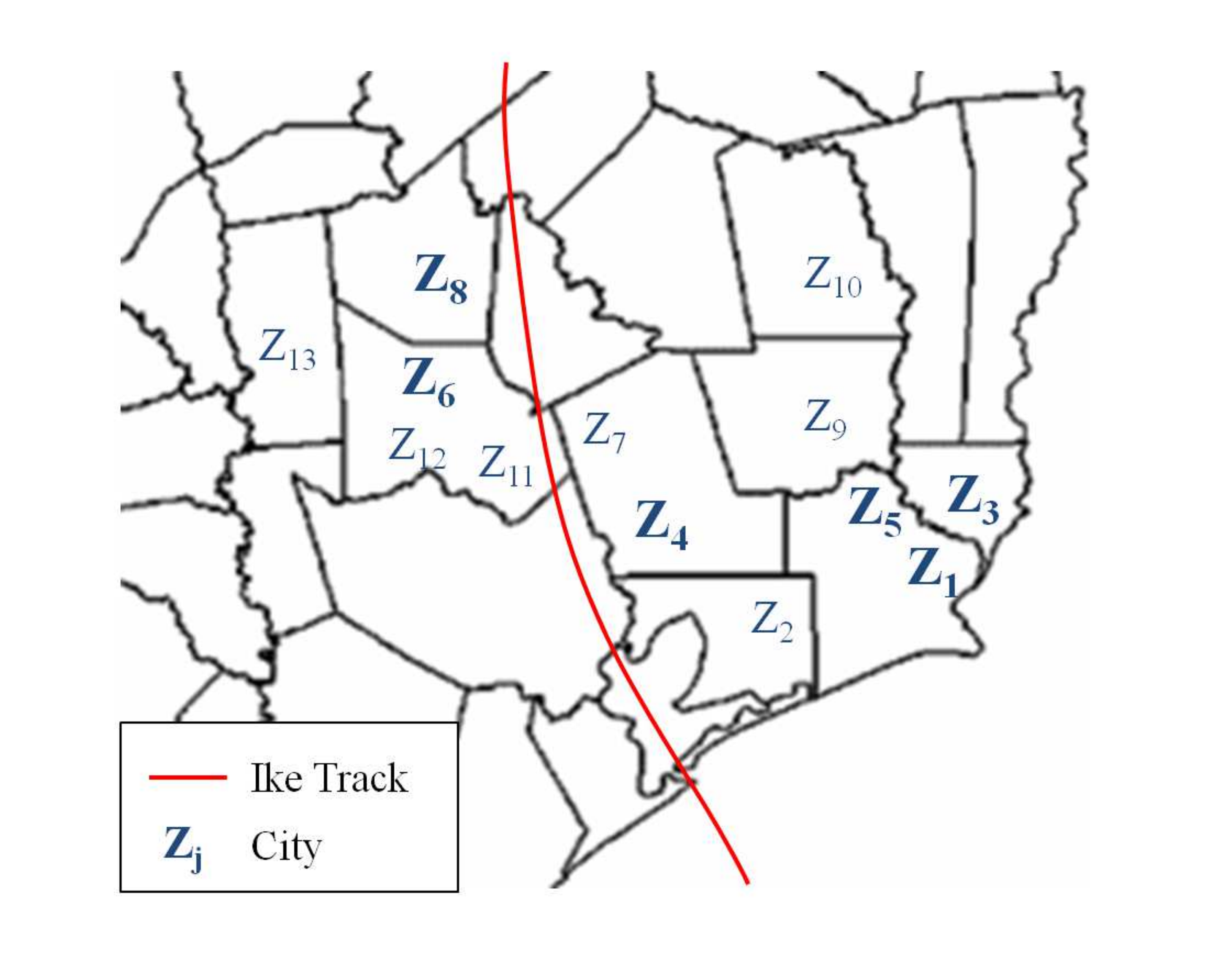}
\caption{Geographical location of the 13 regions (cities).}
\label{fig:GeoCity}
\end{center}
\end{figure}

The parameters of the Weibull mixtures within each interval are learned through maximum likelihood estimation \cite{Duda} from the data. Failure durations obey different distributions for failures occurred at different intervals, showing the non-stationarity. For example, the first duration $\psi_1$ (7 a.m. September 12 to 7 p.m. September 12) is when the network was not yet impacted widely by Hurricane Ike. Three Weibull mixtures are learned from the data, with the shape, the scale and weighting parameters as $(1, 0.71, 0.486)$, $(10.5, 14.4, 0.257)$ and $(10.7, 211.8, 0.257)$. The first two components result in dominating infant recovery, where $74.3\%$ of failures recovered within a day. In contrast, the third duration $\psi_3$ (3 a.m. September 13 to 3 p.m. September 13) is when the large-scale failures continued to occur after the landfall. Two Weibull mixtures are learned from the data. The shape, the scale and weighting parameters are $(5.3, 11.0, 0.323)$ and $(12.4, 112.2, 0.677)$, showing dominating aging recovery. As the result, only $32.2\%$ of failures recovered within a day. The second duration $\psi_2$ (7 p.m. September 12 and 8 a.m. September 13) is around the hurricane landfall, where about a half of the failures occurred experienced infant recovery within a day (see Figure \ref{fig:PDWbl} for the three Weibull mixtures). For $5$ durations overall, the probability of infant recovery within a day changes over time, showing the non-stationary of failure-recovery processes.

We then reconstruct the empirical temporal failure-recovery process $\hat N(t)$ with learned $\hat \lambda_f(t)$ and $\hat \lambda_r(t)$ through Equation \ref{eq:frp}. Figure \ref{fig:FRP} shows the comparisons between $\hat N(t)$ and $N(t)$, the reconstructed and the actual sample paths of the failure-recovery process repectively. The closeness between the two sample pathes shows that the piecewise stationary $g(d|t)$ approximates well the actual failure-and-recovery process.

\begin{figure}
\begin{center}
\includegraphics[width=0.45\textwidth]{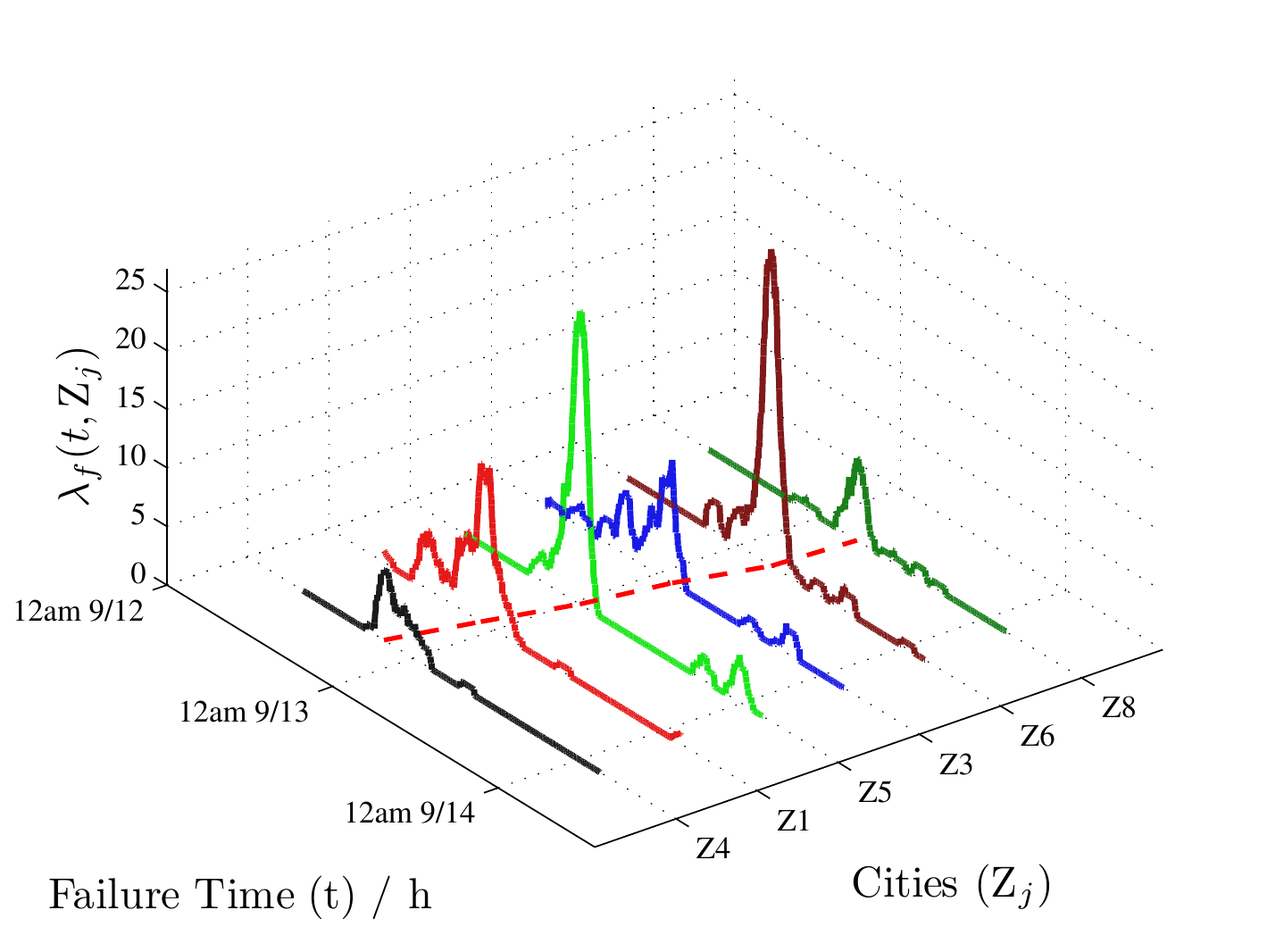}
\caption{Empirical geo-temporal failure rate $\lambda_f$ during Hurricane Ike. Cities are sequenced with respect to the time when the failure rate reached the peak value in each region.}
\label{fig:failurerate}
\end{center}
\end{figure}

\subsection{Geo-Temporal Failure Process}

We now incorporate geo-location variables to learn the geo-temporal non-stationarity. Failure process $N_f(t)$ is a geo-temporal process with multiple attributes $N_f(t, \mathbb{Z}_j)$ from $m$ geographical regions, $1 \le j \le m$. The empirical failure rate functions $\lambda_f(t,\mathbb{Z}_j)$ for $1 \le j \le m$ are estimated using the same algorithm of moving average. The resulting rate vector $\lambda_f(t)$ is multi-variate, consisting of $m$ time-varying functions. Due to the small sample size, there are 6 out of 13 cities shown in Figure \ref{fig:GeoCity}, each of which has sufficient samples ranging from $27$ to $101$. Figure \ref{fig:failurerate} shows the failure rates of the 6 cities. The multi-variate failure rates exhibit the following characteristics:

(a) Temporal non-stationarity: At a given geographical region $ \mathbb{Z}_j$, $\lambda_f(t, \mathbb{Z}_j)$ is a time-varying function similar to the bell-shaped curve obtained for the entire network. Consider $\mathbb{Z}_5$ as an example. The failure rate was low (few than 5 failures) prior to 7 p.m. September 12. Then, the rate increased sharply and reached the maximum value of 25 new failures per hour, at about 1 a.m. September 13. After that, the rate decreased rapidly to few than 5 failures.

(b) Spatial non-stationarity: At a given time $t$, $\lambda_f(t, \mathbb{Z}_j)$ is a spatially-varying function. The peak values of failure rates vary from 1.5 to 27 per hour across the 9 cities. The time when the rate reached the peak value varies between 8 p.m. September 12 to 7 a.m. September 13, and is depicted as a dashed line at the bottom in Figure \ref{fig:failurerate}.

(c) Spatial temporal non-stationarity: The regions are then labeled with respect to the order of failure rates that reached the maximum value in Figure \ref{fig:failurerate}. For example, the failure rate at City $\mathbb{Z}_4$ reached the peak value first, followed by the failure rates at City $\mathbb{Z}_1$ through City $\mathbb{Z}_8$. The figure shows the geo-temporal characteristic that failure rates at different city reached their peak values approximately from the coast to inland. This appears to be consistent to the movement of the hurricane track (Figure \ref{fig:GeoCity}).

\subsection{Geo-Temporal Recovery Process}

To learn the geo-temporal non-stationary recovery, we extend the mixture model (Equation \ref{eq:mix}) to a geo-temporal bivariate mixture, where for $1 \le j \le m$,

\begin{equation}
g(d|t, \mathbb{Z}_j)=\sum_{i=1}^{l(t,\mathbb{Z}_j)}{\rho_i (t,\mathbb{Z}_j) g_{i}(d|t,\mathbb{Z}_j)}.
\label{eq:mix2}
\end{equation}

Again our learning focuses on the 6 cities with sufficient samples. Dependencies of failure durations among cities are not studied in this work because of the small sample size.

\begin{figure}
\begin{center}
\includegraphics[width=0.4\textwidth]{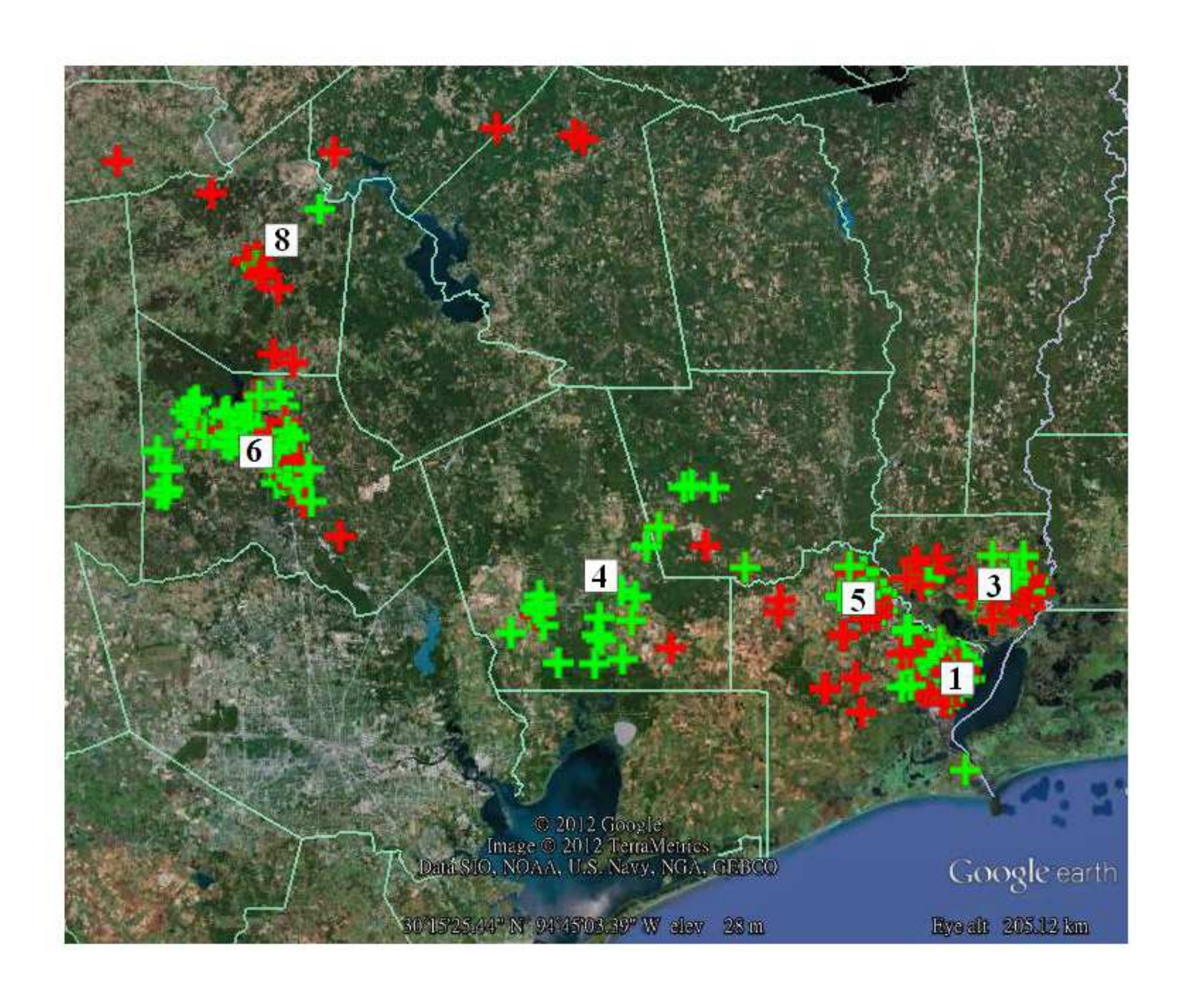}
\caption{Geographical distribution of infant (green) and aging (red) recoveries in the 6 cities: $d_0 = 24$ hours.}
\label{fig:geoIfAg}
\end{center}
\end{figure}

We apply the piecewise homogeneous distribution function in Equation \ref{eq:wblpdf} to each region $\mathbb{Z}_j$,

\begin{equation}
g(d|t \in \psi_i, z \in \mathbb{Z}_j) = \sum_{\zeta=1}^{l_{i,j}} \rho_{\zeta,i,j} g_{\zeta,i,j}(d).
\label{eq:wblpdf_piecewise}
\end{equation}

Here, each component $g_{\zeta,i,j}(d)$ is a Weibull distribution $w(d; \gamma_{\zeta,i,j}, k_{\zeta,i,j})$. Mixture $g(d|t \in \psi_i, z \in \mathbb{Z}_j)$'s and their coefficients vary with respect to not only failure occurrence time $\psi_i$ (temporal non-stationarity) but also geo-locations $\mathbb{Z}_j $'s (spatial non-stationarity).

Applying the maximum likelihood estimation \cite{Duda}, we obtain the estimated parameters of Weibull distributions in the 6 cities. Note that due to the small sample size in some of the regions, the parameters of distributions of failure duration have to be assumed, in our implementation, not varying with failure occurrence time within a region. The probability of infant recoveries is also computed accordingly. Three cities (1, 4, 6) show a similar percentage of infant recovery from $66\%$ to $68\%$ whereas the remaining cities (3, 5, 8) have infant recovery from $40\%$ to $45\%$. Table \ref{tab:wblpar_space} shows the learned model parameters for two example cities. Figure \ref{fig:geoIfAg} shows the geographical distribution of infant and aging recoveries for the 6 cities.

The probability of infant recovery as well as model parameters vary across different geographical regions, showing the spatial non-stationarity of the recovery process. Examining more details, adjacent cities (e.g., 1 and 3) that are close to the coast can exhibit different percentages of infant recovery. Faraway cities (e.g., city 8 which is far in land and city 5 which is close to the coast) can also exhibit a similar percentage of infant recovery. Hence, recovery processes seem to be complex and require further study.

\section{Hurricane Sandy}\label{sec:sandy}

We now learn using real data from another real-life example of large-scale disruptions caused by Hurricane Sandy. This provides an understanding how our model and learning approach can be generalized to other hurricanes.

\begin{figure}
  \centering
  \includegraphics[width=0.45\textwidth]{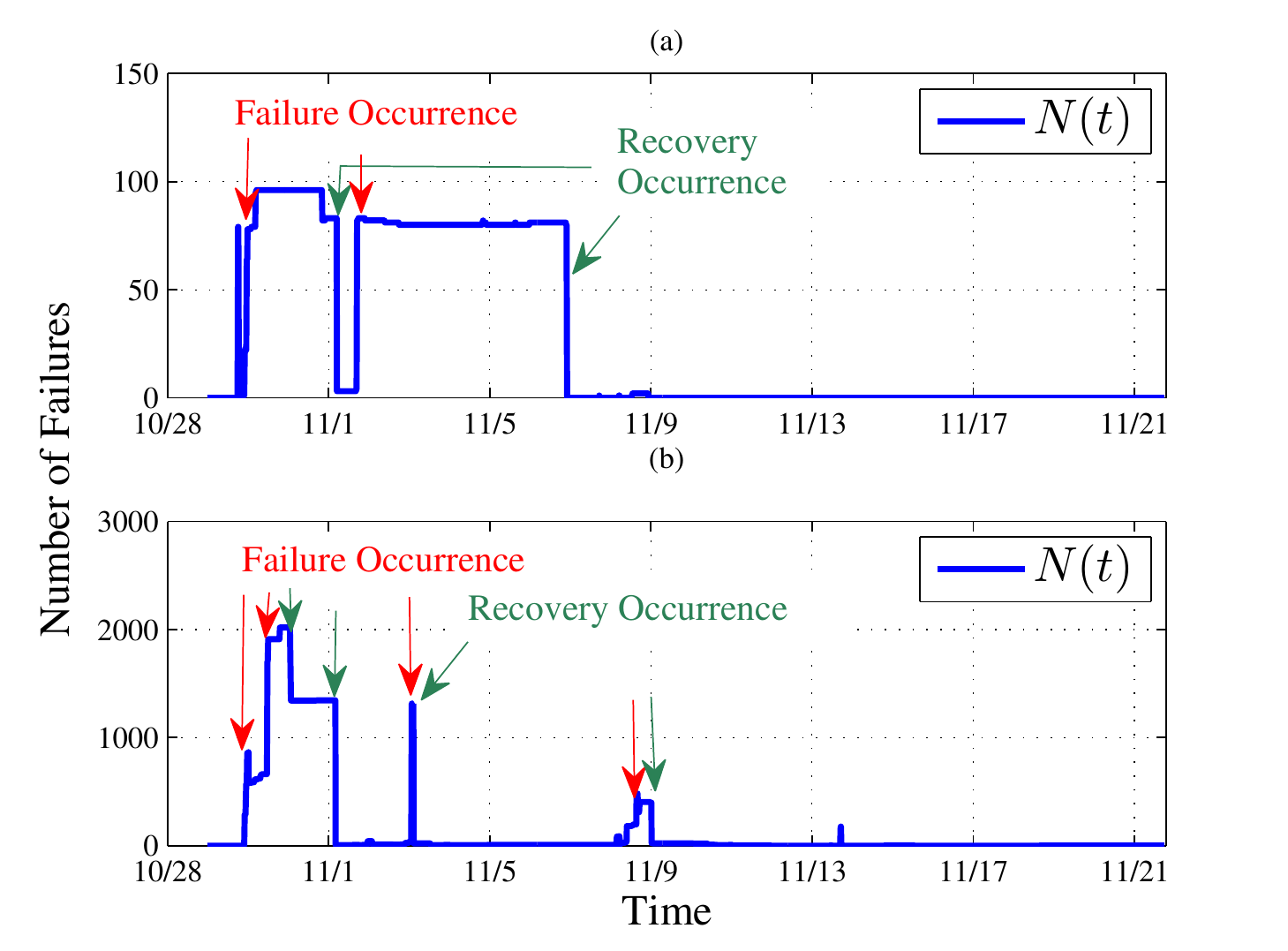}
  \caption{Number of customers without power in two counties of New Jersey.}
  \label{fig:sandy_nt_cnty}
\end{figure}

\subsection{Data}

Hurricane Sandy had a landfall at Northeastern United States on October 28, 2012. Hurricane Sandy resulted more than 6 million customers without electricity for days. The state with the most customers without power was New Jersey, where about 1.98 million customers lost power supplies \cite{Doe_sandy}.

A utility company, reported the number of failures (outages) in more than 10 counties in New Jersey from October 28, 2012 to November 22, 2012. The aggregated number of reported outages is a sample in our data set. Each sample consists of a given geo-location and time $t$ at the scale of $15$ minutes (the reporting interval). The geo-location variable $\mathbb{Z}_j$ corresponds to a county in New Jersey for $1 \le j \le 14$. The data set consists of $2275$ such samples, i.e., $\{N(t,Z_j)\}_{j=1}^{14}$ for time $t$ from October 28 to November 22, 2012. Figure \ref{fig:sandy_fr_rr}(a) plots the data. Note that such aggregated data does not provide accurate occurrence time nor duration of each power failure.

\begin{table}
  \small
  \centering
  \caption{Estimated parameters of distributions of failure durations in 2 cities.} \label{tab:wblpar_space}
  \begin{tabular}{c|c|c|c|c}
      \hline
      $g(d|z\in\mathbb{Z}_1)$ & 1 & 2 & 3 & $P\{d<24\}$ \\
      \hline
      $\rho_{1,\zeta}$    & 0.3478 & 0.3188 & 0.3333 & \\
      $\gamma_{1,\zeta}$  & 0.0045 & 12.1893 & 197.0316 & $66.63\%$ \\
      $k_{1,\zeta}$       & 0.2490 & 2.7891 & 3.7629 & \\
      \hline
      $g(d|z\in\mathbb{Z}_3)$ & 1 & 2 & 3 & $P\{d<24\}$ \\
      \hline
      $\rho_{3,\zeta}$    & 0.3000 & 0.1500 &  0.5500  & \\
      $\gamma_{3,\zeta}$  & 0.0650 & 12.2138 & 129.7408  & $45.37\%$ \\
      $k_{3,\zeta}$       & 0.2897 & 3.9992 & 2.8037 & \\
      \hline
  %    $g_1(d|z\in\mathbb{Z}_4)$ & 1 & 2 & 3 &  $P\{d<24\}$ \\
  %    \hline
  %    $\rho_{4,\zeta}$    &  0.3243 & 0.3514 &  0.3243 & \\
  %    $\gamma_{4,\zeta}$  &  0.0008 & 13.0458 & 156.4058 & $67.62\%$ \\
  %    $k_{4,\zeta}$       &  0.4305 & 3.3091 & 3.2297 & \\
  %    \hline
  %    $g_1(d|z\in\mathbb{Z}_5)$ & 1 & 2 & 3 & $P\{d<24\}$ \\
  %    \hline
  %    $\rho_{5,\zeta}$    & 0.3936 &  0.1489 &  0.4574 & \\
  %    $\gamma_{5,\zeta}$  & 0.0010 & 43.2885 & 191.7912 & $43.35\%$ \\
  %    $k_{5,\zeta}$       & 0.3575 & 1.9824 & 3.9694 & \\
  %    \hline
  %    $g_1(d|z\in\mathbb{Z}_6)$ & 1 & 2 & 3 &  $P\{d<24\}$ \\
  %    \hline
  %    $\rho_{6,\zeta}$    & 0.1900 & 0.4700 & 0.3400 & \\
  %    $\gamma_{6,\zeta}$  & 0.0096 & 13.2207 & 151.7052 & $65.96\%$ \\
  %    $k_{6,\zeta}$       & 0.2228 & 4.4916 & 4.0586 & \\
  %    \hline
  %    $g_1(d|z\in\mathbb{Z}_8)$ & 1 & 2 & 3 &  $P\{d<24\}$ \\
  %    \hline
  %    $\rho_{8,\zeta}$    & 0.3333 & 0.1852 & 0.4815 & \\
  %    $\gamma_{8,\zeta}$  & 0.0356 & 34.4387 & 155.3905 & $40.67\%$ \\
  %    $k_{8,\zeta}$       &  0.2146 & 1.6564 & 3.3677 & \\
  %    \hline
  \end{tabular}
\end{table}

\subsection{Empirical Failure Process}

Learning now begins with the aggregated number of failures $N(t, Z_j)$ for $1 \le j \le 14$, from which failure- and recovery- rates are estimated accordingly. This is a reverse process to learning from detailed failure data in Hurricane Ike.

To learn the failure rate, we recall that $\lambda_f(t) = \frac{d}{dt} E[N_f(t)]$ from Equation \ref{eq:fp}, and $\lambda_f(t)-\lambda_r(t) = \frac{d}{dt} E[N(t)]$ from Equation \ref{eq:frp}. This suggests that a lower bound $\hat \lambda_{fl}(t)$ on the failure rate can be estimated from the aggregate number of failures at time $t$ as
\begin{equation}
\hat \lambda_{fl}(t, \mathbb{Z}_j) = \frac {d}{dt} N(t, \mathbb{Z}_j), \qquad \text{if} \quad t=t^\ast,
\label{eq:sandy_fr_fit}
\end{equation}
where $t^\ast$ is a time epoch when $N(t^*, \mathbb{Z}_j)$ increases.

To determine how to obtain such an estimate, we examine characteristics of raw (time series) data $N(t,Z_j)$ at the county level. Figure \ref{fig:sandy_nt_cnty} shows two examples of the number of aggregated failures $N(t, \mathbb{Z}_j)$ at two different counties in New Jersey. $N(t, \mathbb{Z}_j)$ shows sharp increases and sharp decreases. A sharp increase occurs when the failure rate exceeds the recovery rate whereas a sharp decrease happens when recovery rate exceeds the failure rate. Hence, a change point in $N(t, \mathbb{Z}_j)$ can be used to identify a lower bound for either a failure rate or a recovery rate. In addition, a sharp increase/decrease indicates a salient rather than noisy change point, where a lower bound can be obtained accurately.

\begin{figure}
  \centering
  \includegraphics[width=0.45\textwidth]{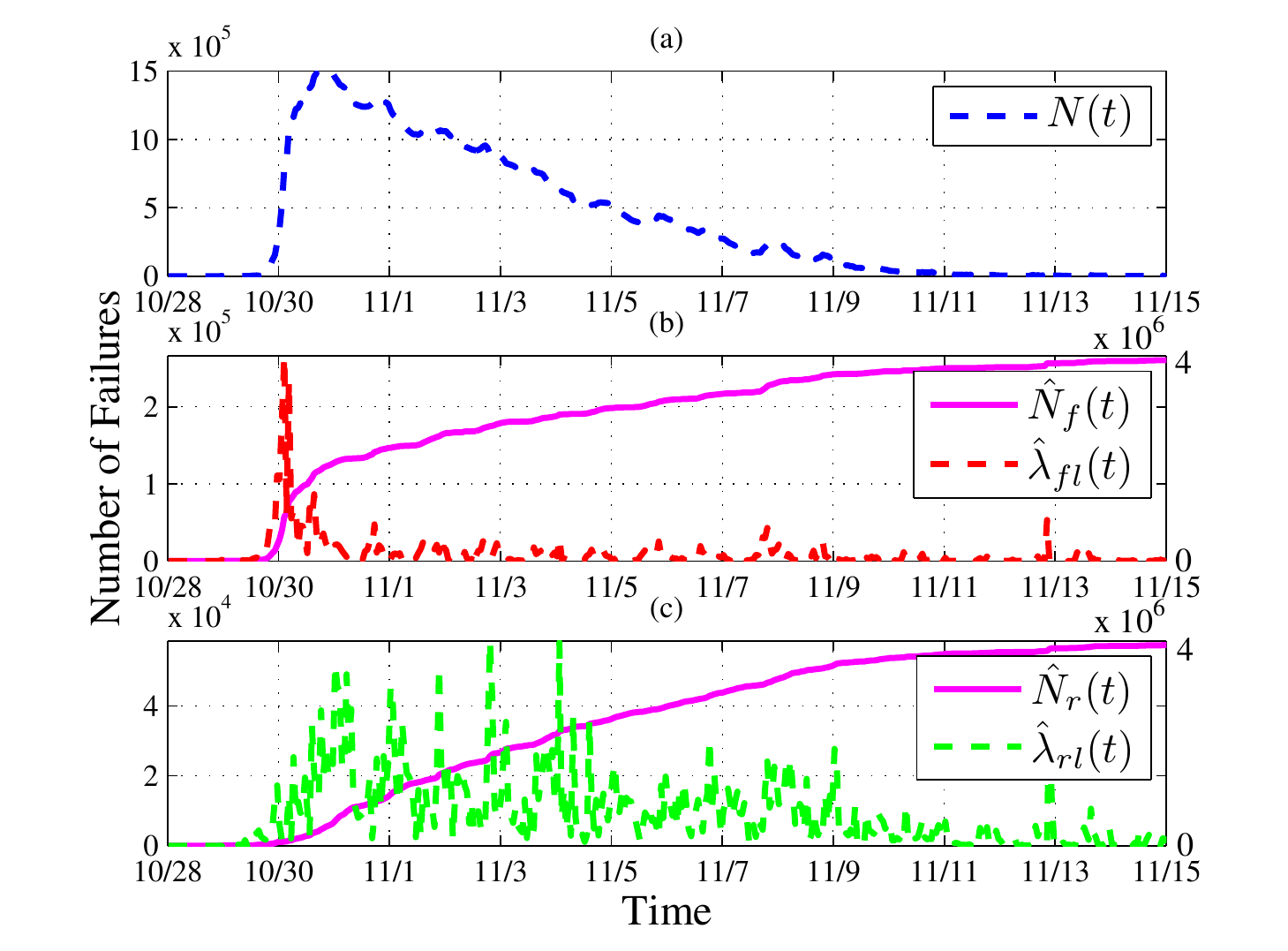}
  \caption{Failure process and recovery process from Hurricane Sandy: (a) $N(t)$, (b) $\hat \lambda_{fl}(t)$, (c) $\hat \lambda_{rl}(t)$.}
  \label{fig:sandy_fr_rr}
\end{figure}

We first obtain the positive increments from $N(t, Z_j)$ for each region $Z_j$ using Equation \ref{eq:sandy_fr_fit}. We then aggregate the increments over the $14$ regions to obtain a lower bound $\hat \lambda_{fl}(t)$ for the failure rate of the utility network. $\hat N_f(t)$, the estimated lower bound on the number of failures up to time $t$, can then be obtained by integrating $\hat \lambda_{fl}(t)$, which is shown in Figure \ref{fig:sandy_fr_rr}(b) \footnote{ EST is used for plots in regard to Hurricane Sandy.}.

\subsection{Empirical Recovery Process}

To learn the empirical recovery rate, we apply Equation \ref{eq:sandy_fr_fit} except that $t^\ast$ corresponds to the time epoch of a decrease in the number of failures. Figure \ref{fig:sandy_fr_rr}(c) shows an estimated lower bound $\hat \lambda_{rl} (t)$ for recovery rate and the cumulative number of recoveries $\hat N_r(t)$ respectively.

\begin{figure}
  \centering
  \includegraphics[width=0.45\textwidth]{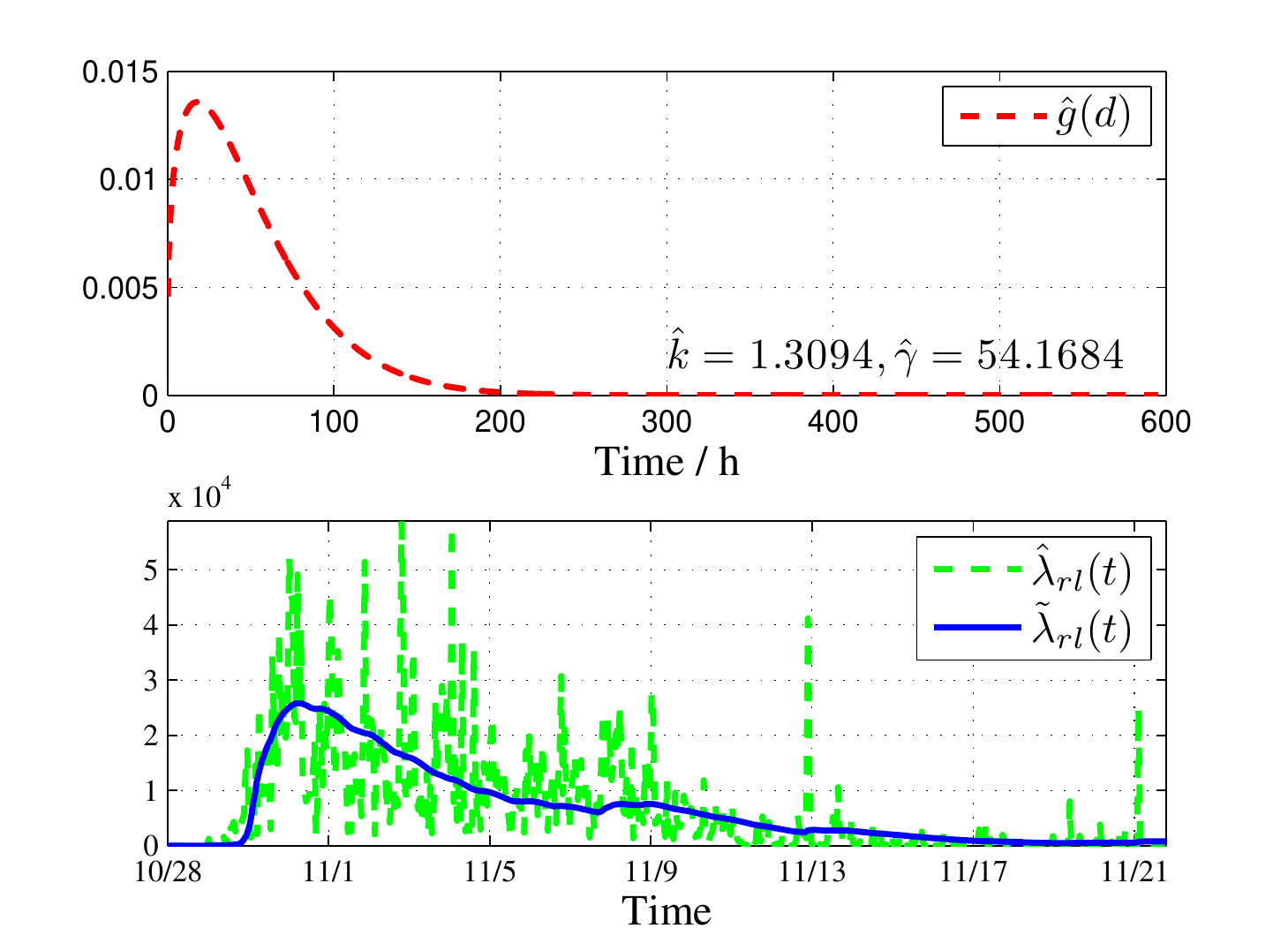}
  \caption{Weibull distribution for failure duration $\hat g(d)$.}
  \label{fig:sandy_gdt}
\end{figure}

Since the aggregated data from Hurricane Sandy does not contain detailed recovery time for each failure, it is impossible to learn the time-varying distribution of failure duration $g(d|t)$. Nevertheless, the aggregated data can be used to estimate a stationary distribution of recovery time, i.e., $g(d)$. As the detailed information on failure duration is not available from the data, we consider a simple distribution with one Weibull mixture $g(d;\gamma, k)$. Applying discrete samples to Theorem \ref{thm:rp}, reconstructed recovery rate $\tilde \lambda_{rl}(t)$ can be related with $g(d;\gamma, k)$ and $\hat \lambda_{fl}(t)$ as
\begin{equation}
\tilde \lambda_{rl}(i \cdot \delta) \approx \sum_{j=0}^{i} g(i \cdot \delta - j \cdot \delta) \hat \lambda_{fl}(j \cdot \delta) \delta,
\end{equation}
where $\delta = 15$ minuets is the step size, and $i \delta$ is the discrete time. Weibull parameters $\gamma$ and $k$ are then estimated to minimize the estimation error $||\tilde \lambda_{rl}(t) - \hat \lambda_{rl}(t)||^2$. Figure \ref{fig:sandy_gdt} shows the estimated Weibull distribution, where the shape parameter $\hat k = 1.3094$ and the scale parameter $\hat \gamma = 54.1684$. The resulting stationary distribution of failure durations is then used to reconstruct a lower bound for the recovery rate. Figure \ref{fig:sandy_gdt} shows the estimated $\hat \lambda_{rl}(t)$ from the data set and the reconstructed $\tilde \lambda_{rl}(t)$. Reconstructed $\tilde \lambda_{rl}(t)$ thus provides a profile on how the recovery varies with time.

\section{Findings and Discussions}\label{sec:discuss}

\subsection{Findings}

Learning from Hurricane Ike and Hurricane Sandy results in the following findings.

\subsubsection{Failure process}

Failure rates are time-varying for both Hurricane Ike and Hurricane Sandy. The corresponding failure processes are non-stationary in time and geo-graphical regions. However, the failure rates exhibit different characteristics at the county level for Hurricane Ike and Hurricane Sandy: The failure rates for Hurricane Ike appear to vary gradually. However, the failure rates for Hurricane Sandy exhibit sharp changes, showing that failures occurred in groups \footnote{The cause shall be sought for when more detailed data becomes available.}. When aggregated over geographical regions, failure rates for both hurricanes exhibit similar characteristics, i.e., first rapidly increasing and then decreasing. More quantitative study is needed to further compare the failure processes for different hurricanes at different spatial scales.

\subsubsection{Recovery process}

Learned recovery rates from Hurricane Ike and Hurricane Sandy are both time-varying. For Hurricane Ike, the learned probability distributions of failure durations exhibit non-stationarity in time and geo-locations, i.e., depend on when failures occur. Such distributions constitute both infant and aging recovery, as shown in Table \ref{tab:wblpar_space} and Figure \ref{fig:geoIfAg}. The degree of infant recovery, however, is different at different cities. Three out of the six chosen cities recovered more rapidly then the rest. Failures with infant and aging recoveries are also inter-leaving in geo-locations.

The recovery for the provider network from Hurricane Sandy shows a nearly steady rate of 7000 recoveries per hour. In addition, the estimated Weibull distribution of the failure duration exhibits stronger aging recovery than infant recovery. A lack of infant recovery for this utility provider may indicate that power distribution networks suffered virulent disruptions during Hurricane Sandy. The recovery can thus be difficult. Yet, detailed rather than aggregated failure data is needed for accurately estimating distributions of failure durations.

Note that failures and recoveries can occur simultaneously within a $15$ minute interval. That is why the amount of increase in $N(t, Z_j)$ is a lower bound of the actual failure rate $\lambda_f(t, Z_j)$. When the number of failures increased rapidly, e.g., from October 28 to October 31, recovery appeared to be minor. When the hurricane passed the area after October 31, recovery dominated. This is shown by the lower bounds of the failure- and the recovery-rate in Figure \ref{fig:sandy_fr_rr} and \ref{fig:sandy_gdt}.

\subsection{Discussions}

The type of available data is important for learning non-stationary behaviors of power distribution in response to external disruptions. The accurate failure data from Hurricane Ike characterizes an entire life cycle of failure and recovery processes. Data from Hurricane Sandy is aggregated and thus lack of exact information on individual failure occurrence and duration. Hence, learning is to infer failure- and recovery-processes, which is a reverse process to that for Hurricane Ike. The 15-minute sampling interval seems to be sufficient for estimating the lower bounds of failure- and recovery-rates from Hurricane Sandy. The aggregated data is insufficient for characterizing a non-stationary distribution of failure duration but can be used to learn a stationary distribution as an approximation.

To deal with the small sample size, a rule of thumb is used where training samples should be several times more than parameters \cite{Duda}. For Hurricane Ike, 20 or more samples seem to be sufficient for estimating temporal characteristics of failure- and recovery-rates but insufficient when the spatial non-stationarity is studied. This suggests that the algorithm need to be enhanced, e.g., to identify spatial scales appropriate for aggregation.

Our model assumes an underlying radial topology, where failures can be considered as independent increments at large temporal spatial scales (minutes, cities). Detailed network configuration is yet to be included in our model. For example, topology and power flows \cite{Wu89}\cite{Zhao13} are two possible characteristics to be included for failures and recoveries. Failure- and recovery-process at a small time scale of sub-seconds then need to be considered accordingly. A challenge is much increased complexity and in-network measurements at temporal spatial scales.

\section{Conclusion}\label{sec:Conclusion}

This work shows that non-stationary geo-temporal random processes naturally model large-scale failure and recovery of power distribution induced by hurricanes. In particular, multivariate geo-location based $GI(t)/G(t)/\infty$ queues provide such non-stationary failure- and recovery processes. The non-stationary failure and recovery can be completely characterized to the expected values by time-varying failure rate and probability distribution of recovery time across geo-graphical regions.

Real data from two hurricanes have been used to learn failure and recovery processes. Learning detailed failure data from Hurricane Ike reveals that the failure process across different geographical regions follows a similar trend to that of the hurricane. However, the failure- and recovery-processes exhibit different infant and aging recovery across geo-graphical regions. Learning aggregated data from an impact area by Hurricane Sandy shows that our model can infer failure- and recovery rates using aggregated data. The failure rates have more significant discrete components for Hurricane Sandy than for Hurricane Ike at geographical regions. The recovery process is dominated by aging recovery for one utility network from Hurricane Sandy but consists of a significant component of infant recovery for another utility from Hurricane Ike. This shows that $GI(t)/G(t)/\infty$ model is indeed needed for general failure- and recovery-processes in dynamic queues. Note that these findings are for power distribution through open rather than underground networks.

These findings call for subsequent research on how distributed power distribution are impacted by external disturbances. For example, power failures and recoveries are yet to be studied at all impact areas for Hurricane Sandy. Spatial temporal dependencies among power distribution networks at different geographical regions need to be studied explicitly. This requires combining detailed configurations of power distribution with the dynamic model. These studies shall provide further understanding on how to enhance the distributed power infrastructure.

\section{Acknowledgement}

The authors would like to thank Chris Kung, Jae Won Choi, Daniel Burnham, Xinyu Dai and Michael Perez for data processing, Amanda Cox for providing parts of the data and helpful discussions, Anthony Kuh for helpful discussions on distribution networks, anonymous reviewers for valuable comments, and Associate Editors for helpful suggestions. Support from National Science Foundation (ECCS 0952785) is gratefully acknowledged.

\bibliographystyle{IEEEtran}
\bibliography{tnnref}

\end{document}